\documentclass[11pt]{article}
\pdfoutput=1
\usepackage{amsmath}
\usepackage{amssymb}
\usepackage{amsfonts}
\usepackage{amscd}
\usepackage{accents}
\usepackage{epsfig}
\usepackage[utf8]{inputenc}
\usepackage{graphicx,amssymb,amsmath,wasysym}

\usepackage{bm}
\usepackage{color}
\usepackage{hhline}
\usepackage[pdfborder={0 0 0}]{hyperref}

\date{\today}
\newcommand{\bmat}{\left(\begin{array}}
\newcommand{\emat}{\end{array}\right)}
\newcommand{\be}{\begin{equation}}
\newcommand{\ee}{\end{equation}}
\newcommand{\ba}{\begin{eqnarray}}
\newcommand{\ea}{\end{eqnarray}}

\def\lsim{\raise0.3ex\hbox{$\;<$\kern-0.75em\raise-1.1ex\hbox{$\sim\;$}}}
\def\gsim{\raise0.3ex\hbox{$\;>$\kern-0.75em\raise-1.1ex\hbox{$\sim\;$}}}

\def\be{\beta}

\usepackage{graphicx}
\usepackage[tight]{subfigure}

\def\bal#1\eal{\begin{align}#1\end{align}} 
 \newcommand{\bea}{\begin{eqnarray}}
\newcommand{\eea}{\end{eqnarray}}
\newcommand{\besub}{\begin{subequations}}
\newcommand{\eesub}{\end{subequations}}
 
\newcommand{\bi}{\begin{itemize}}
\newcommand{\ei}{\end{itemize}}

\newcommand{\Mcal}{{\cal M}}

\newcommand{\lh}{\ensuremath{\lambda_h}}
\newcommand{\ls}{\ensuremath{\lambda_s}}
\newcommand{\lhs}{\ensuremath{\lambda_{hs}}}

\begin{document}

\begin{center}
 {\bf \LARGE Neutrino dark matter  and the Higgs portal: improved freeze--in analysis } \\
\end{center}

\vskip0.5in

\begin{center}
 { \large Valentina De Romeri$^{a,}$\footnote{email: {\tt deromeri@ific.uv.es}}, Dimitrios Karamitros$^{b,}$\footnote{email: {\tt Dimitrios.Karamitros@ncbj.gov.pl}}, Oleg Lebedev$^{c,}$\footnote{email: {\tt oleg.lebedev@helsinki.fi}}, Takashi Toma$^{d,e,}$\footnote{email: {\tt toma@staff.kanazawa-u.ac.jp}}} 
\vspace{0.8cm}\\
{\it {\sl \small  
 $^a$ Institut de F\'{i}sica Corpuscular CSIC/Universitat de Val\`{e}ncia, Parc Cient\'ific de Paterna\\
 C/ Catedr\'atico Jos\'e Beltr\'an, 2 E-46980 Paterna (Valencia) - Spain\\[3mm]

 $^b$ National Centre for Nuclear Research, ul. Pasteura 7, 02-093 Warsaw, Poland \\[3mm]

 $^c$Department of Physics, University of Helsinki, Gustaf H\"allstr\"omin katu 2a, Helsinki, Finland\\[3mm]

 $^d$Department of Physics, McGill University, 3600 Rue University, Montr\'{e}al, Qu\'{e}bec H3A 2T8, Canada 

$^e$ Institute of Liberal Arts and Science, Kanazawa University, Kakuma-machi, Kanazawa 920-1192, Japan}}
\end{center}

\vskip0.5in

 { ABSTRACT: Sterile neutrinos are one of the leading dark matter candidates. Their masses may originate from a vacuum expectation value of a scalar  
 field.  If the sterile neutrino couplings are very small and their direct coupling to the inflaton is forbidden by the lepton number symmetry, the leading dark matter production mechanism is the freeze--in scenario. We study this possibility in the neutrino mass range up to 1 GeV,
  taking into account  relativistic production rates based on the Bose--Einstein statistics,  thermal masses and  phase transition effects. 
  The specifics of the production mechanism and the dominant mode depend on the relation between the scalar and sterile neutrino masses as well as on whether or not the scalar is thermalized. We find that the observed dark matter abundance can be produced in all of the cases considered. We also revisit the freeze--in production of a Higgs portal scalar, pointing out the importance of a fusion mode,  
  as well as the thermalization constraints.  } \\

\vskip0.5in

\newpage

\tableofcontents

\newpage

\section{Introduction}

One of the  outstanding mysteries of the Universe is the nature of dark matter (DM). 
An attractive minimal option is provided by sterile neutrinos, whose existence is strongly suggested by the observed neutrino masses.
The smallness of the latter can elegantly  be explained by the \textit{seesaw} mechanism~\cite{Minkowski:1977sc,GellMann:1980vs,Yanagida:1979as,Mohapatra:1979ia,Schechter:1980gr,Lazarides:1980nt}.
When the active--sterile neutrino mixing is sufficiently small, 
the lightest sterile neutrino can  be very long--lived and play the role of DM.
   In the simplest scenario proposed by Dodelson and Widrow~\cite{Dodelson:1993je}, such neutrinos can be produced via mixing with the active  neutrinos in a thermal bath of the Standard Model (SM) particles, although the sterile neutrinos do not thermalize themselves. This minimal option now appears to be in conflict with a number of observations~\cite{Boyarsky:2005us,Seljak:2006qw,Boyarsky:2006ag,Boyarsky:2006fg,Boyarsky:2007ay,Boyarsky:2007ge,Yuksel:2007xh,Boyarsky:2008xj,Ackermann:2015lka,Perez:2016tcq} (see e.g.~\cite{Adhikari:2016bei} for a review).

 Other production mechanisms   have been explored in the literature. 
 Primordial lepton asymmetry could generate the active--sterile transitions via the Shi and Fuller mechanism~\cite{Shi:1998km,Abazajian:2001nj}. 
 This option has been studied  extensively  in the context of the neutrino Minimal Standard Model ($\nu$MSM)~\cite{Asaka:2005an,Asaka:2006nq,Canetti:2012vf,Canetti:2012kh,Boyarsky:2009ix}. 
 Alternatively, the relic population of sterile neutrinos may   be generated via  decay of a heavier particle like the inflaton~\cite{Shaposhnikov:2006xi,Bezrukov:2009yw}, the radion~\cite{Kadota:2007mv} or a general scalar singlet~\cite{Kusenko:2006rh,Petraki:2007gq,Merle:2013wta,Adulpravitchai:2014xna,Frigerio:2014ifa,Merle:2015oja}. Other possibilities include sterile neutrino  production through  pion decays~\cite{Lello:2014yha}, heavy pseudo-Dirac neutrinos~\cite{Abada:2014zra},   interactions of light vector bosons~\cite{Boyanovsky:2008nc,Shuve:2014doa}, via an axion--like field~\cite{Berlin:2016bdv} and parametric resonance \cite{Bezrukov:2019mak}.

 An interesting possibility,  which we explore in detail, is  
  to generate  the observed relic abundance of the sterile neutrino DM through the \textit{freeze--in} mechanism~\cite{McDonald:2001vt,Hall:2009bx}. 
  This scenario requires a tiny coupling  and a negligible initial DM abundance. The correct   relic density is then gradually built up via this feeble coupling along the evolution of the Universe. 
  A successful realization of this mechanism involves an extra scalar field, whose decay produces sterile neutrinos~\cite{Kusenko:2006rh,Petraki:2007gq}. 
  The  vacuum expectation value (VEV) of such a  field can be responsible for the Majorana neutrino masses~\cite{Chikashige:1980ht,Chikashige:1980ui}.
  For a small enough coupling, the freeze--in mechanism is at work and the correct DM density can be produced.
  
  In our work, we perform an in--depth analysis of the freeze--in production of sterile neutrinos in the mass range up to 1 GeV, taking into account different possible production regimes, relativistic reaction rates with the Bose--Einstein distribution function, thermal masses and main effects of the phase transitions. 
  Previous studies have mainly focused on the keV mass range~\cite{Kusenko:2006rh,Petraki:2007gq,Shakya:2015xnx,Merle:2015oja,Drewes:2015eoa}, in which case the active--sterile mixing angle is required to be below $10^{-5}$ or so. In our case, the requisite mixing must be much smaller calling for a symmetry justification. The required symmetry  can be identified with  the neutrino parity which acts  on the lightest sterile neutrino only. 
  
  We find that the neutrino production often takes place in the relativistic regime, in which case the  Bose--Einstein distribution should be used for the initial state scalars.
  This differs from many previous studies which have resorted to the non--relativistic Maxwell-Boltzmann approximation. In order to take quantum statistics into account, 
  we follow the approach of \cite{Arcadi:2019oxh},\cite{Lebedev:2019ton} and extend it to asymmetric reactions. The resulting rate enhancement depends strongly on the thermal masses and ranges from ${\cal O}(1)$ to two orders of magnitude in the vicinity of the 2d order phase transition.
  
  We also take into account the most important  effects  of the phase transitions. First of all, these affect the presence or absence of certain couplings which depend on scalar VEVs.
  In addition, the mass change at the phase transition can facilitate particle production.
  
  The DM production modes depend on whether or not the  scalar is thermalized. It couples to the SM via the Higgs portal~\cite{Silveira:1985rk,Schabinger:2005ei,Patt:2006fw}.  Then, its thermalization depends on  the Higgs portal coupling and the maximal temperature.
  To this end, we revisit the scalar production through the Higgs portal couplings and the consequent thermalization constraints. We find, in particular, that the $2\rightarrow 1$ 
  reaction (fusion) plays an important role in this analysis.
  
  In this work, we are mainly interested in reproducing the correct DM relic abundance. To this end, we solve the Boltzmann equation for the number density rather than the 
  momentum distribution function (unlike e.g. \cite{Merle:2015oja}). We reserve the latter for future work.
   
  The paper is organized as follows. In Section~\ref{model}, we introduce our model, discuss the leading  thermal corrections and compile  the current constraints on  sterile neutrino DM. In Section~\ref{asym}, we generalize the relativistic reaction rates  of \cite{Arcadi:2019oxh},\cite{Lebedev:2019ton} to asymmetric reactions. Thermalization constraints are discussed in
  Section~\ref{thermalization}. Our main results are presented in Sections~\ref{secI} and~\ref{secII}. We conclude in Section~\ref{conclusion}.

\section{The model}
\label{model}

  In this work, we focus on a  simple set--up:  the SM is extended by a $real$ singlet  $S$ and some number of right--handed (sterile) neutrinos $\nu_{R_i}$.\footnote{Their number can be significantly larger than 3 as motivated by string theory \cite{Buchmuller:2007zd}. Here, a sterile neutrino is defined as a fermion that has a Yukawa coupling with the SM neutrinos as well as a Majorana mass term.    } The lightest of them is assumed  to constitute long--lived dark matter. 
        
  We assume that the Majorana masses are produced entirely by the singlet scalar VEV. This can be implemented through a lepton--number discrete symmetry forbidding the bare mass:
  \begin{equation}
  S \rightarrow -S ~~,~~ \nu_i \rightarrow i \nu_i \;.
  \end{equation}
  The relevant Lagrangian terms are then
  \begin{equation}
  -\Delta {\cal L} = {1\over 4} \lambda_s S^4 +{1\over 2} \mu_s^2 S^2 + {1\over 2 }  \lambda_{hs} \vert H\vert^2 S^2 + {1\over 2} \lambda_{ij} S \; \nu_{R_i} \nu_{R_j} + y_{ij} H^c \; \bar l_i  \nu_{R_j}\;. 
  \end{equation}
  The above symmetry results in 2 useful  properties:
  \begin{itemize}
\item  no inflaton coupling to $\nu_{R_i} \nu_{R_j}$ is allowed (assuming that the inflaton carries no lepton charge). Otherwise, inflaton decay would normally dominate the neutrino production.
  \item diagonalizing the neutrino mass matrix diagonalizes the $S$ coupling to neutrinos (neglecting the Dirac contributions). As a result, there is  no flavor change and a heavier $\nu$ cannot produce a lighter $\nu$ via its decay with $S$--emission. Thus, we can focus on direct freeze--in production of the lightest $\nu$.
  \end{itemize}
  
  Let us denote the lightest mostly--sterile neutrino $\nu$ and its coupling to $S$ $\lambda$:
   \begin{equation}
    -\Delta {\cal L} = {1\over 2} \lambda\; S \; \nu \nu \;. 
  \end{equation}
  Its mass is then $M = \lambda \langle S \rangle$ neglecting the Dirac mass contribution. Throughout this paper we assume that the relevant Yukawa couplings are very small such that
  the usual Higgs decay does not produce a significant amount of dark matter. The resulting active--sterile mixing angle $\Theta \sim y \langle H \rangle / \lambda \langle S \rangle \ll 10^{-5}$ is also very small.

 \subsection{The scalar sector}
  
 The scalar sector of the model includes the Higgs field  $H$ and the real scalar $S$. The potential invariant under the $S \rightarrow -S$ symmetry is given by
  \bal \label{pot}
V(h,s)=\frac{\lh}{4} h^4 + \frac{\lhs}{4} h^2 S^2 + \frac{\ls}{4} S^4 + \frac 12 \; \mu_h^2 \, h^2 + \frac 12 \mu_s^2 \, S^2 \,.
\eal
Here we use the unitary gauge  $H=(0,h/\sqrt{2})^T$.
Both $H$ and $S$ must develop non--zero VEVs $v$ and $u$, respectively. These are given by  
\besub
\bal
v^2&=\frac{2 \lhs \mu_s^2 - 4 \ls \mu_h^2}{4 \lh \ls -\lhs^2}  
\\
u^2&=\frac{2 \lhs \mu_h^2 - 4 \lh \mu_s^2}{4 \lh \ls -\lhs^2}   \;.
\eal
\eesub
The mass matrix at this point is
\bal
\Mcal^2= \left( \begin{array}{cc}
2 \lh v^2 & \lhs v u \\
\lhs v u & 2 \ls u^2
\end{array} \right) \,.
\eal
Since the couplings are real and we require $v^2>0,u^2>0$, the mass matrix $\Mcal^2$ is positive definite if and only if 
\begin{equation}
\lh>{\lhs^2 \over 4 \ls} ~~,~~ \ls>0 ~. \label{condition}
\end{equation}
$\Mcal^2$ can be diagonalized by the orthogonal transformation
\bal
O^T \Mcal^2 O = \textrm{diag}(m_1^2,m_2^2) ~,
\eal
where
\bal
O&= \left( \begin{array}{cc}
\cos \theta & \sin \theta \\
- \sin \theta  & \cos \theta
\end{array} \right) \;
\eal
and the angle $\theta$ satisfies
\bal
\tan 2 \theta  &= \frac{ \lhs v u}{\ls u^2-\lh v^2} \,. \label{tantwotheta}
\eal
The mass squared eigenvalues are given by
\bal
m_{1,2}^2 = \lh v^2+\ls u^2\mp \frac{ \ls u^2 - \lh v^2}{\cos 2 \theta} \,.
\label{evalues}
\eal
 The above equation implies $\textrm{sign}(m_2^2-m_1^2)=\textrm{sign}(\cos 2 \theta) \,\textrm{sign}(\ls u^2 - \lh v^2)$.
We will  primarily be interested in the small mixing case. (E.g., for a light singlet, the meson decay and LEP constraints on the mixing angle are very strong~\cite{Falkowski:2015iwa}.) Thus,
it is convenient to employ the small angle approximation $\sin \theta \ll 1$ and neglect the $\theta^2$--terms.
In this case,  the eigenvalues can be relabelled according to the state composition and 
satisfy
\begin{equation}
m_h^2 \simeq 2 \lambda_h v^2 ~~,~~m_s^2 \simeq 2 \lambda_s u^2 \;. 
\label{appr}
\end{equation}
The mixing angle can then be expressed as 
 \begin{equation}
\theta \simeq {\lambda_{hs} \over \sqrt{4 \lambda_h \lambda_s}} ~ {m_s m_h \over m_s^2 - m_h^2} \;.
\end{equation}
This form is convenient since stability considerations bound the first factor  by 1.
Clearly, for $m_s$ close to $m_h$ our approximation fails. When $m_h$ and $m_s$ are substantially different, the mixing angle is bounded by
 $\vert\theta\vert < m_s/m_h \;, m_h/m_s$.
 In most of our parameter space, the mixing angle is indeed very small.

  In what follows, the sign of $\theta$ is unimportant, so will denote by $\theta$ the magnitude of the mixing angle.

  \subsubsection{Thermal corrections}
  
  At high temperature, the scalar potential gets modified by the thermal corrections. The main effect is captured by the thermal masses which amount to the replacements
  \begin{equation}
\mu_h^2 \rightarrow \mu_h^2 + c_h T^2 ~~,~~ \mu_s^2 \rightarrow \mu_s^2 + c_s T^2 \;,
\label{ther-mass}
\end{equation}
where
\begin{eqnarray}
&& c_h \simeq {3 \over 16} g^2 + {1\over 16} g^{\prime 2} + {1\over 4} y_t^2 + {1\over 2} \lambda_h \;, \nonumber\\
&& c_s= {1\over 4} \lambda_s + {1\over 6} \lambda_{hs}\;.  
\end{eqnarray}
Here $g, g^\prime$ are the SM gauge couplings and $y_t$ is the top--quark Yukawa coupling.
At high $T$, the minimum of the potential is at  $v=u=0$. The transition to non-zero VEVs takes place at the critical  temperatures: 
$v=0 \rightarrow v\not=0$ at $T_c^v$ and $u=0 \rightarrow u\not=0$ at $T_c^u$. 
The dynamics of the transition is somewhat complicated and proceeds in steps: at the first stage, one of the VEVs stays zero and the other becomes non--zero, while at the second stage both of the fields attain non--zero VEVs. On the other hand, we find that 
 the neutrino DM production depends on the critical temperature rather weakly. Therefore, it suffices for our purposes to approximate the critical temperature by that of the first stage, 
 $T_c^2= \vert \mu_i^2 \vert /c_i$.
The critical temperatures  can then be expressed in terms of the physical masses and the couplings for 
$\sin\theta \ll 1$:
\begin{eqnarray}
&& T_c^v = {  \vert \mu_h \vert \over  \sqrt{c_h}  } \;, \nonumber\\
&&  T_c^u = {  \vert \mu_s \vert \over  \sqrt{c_s}  }      \;, \nonumber\\
&& -\mu_h^2 = {\lambda_{hs}\over  4 \lambda_s}\; m_s^2 +{1\over 2} m_h^2 \;, \nonumber\\
 &&     -\mu_s^2 = {\lambda_{hs}\over  4 \lambda_h}\; m_h^2 +{1\over 2} m_s^2 \;.
\end{eqnarray}
 
It is important to include the thermal masses (\ref{ther-mass}) in the calculation of the reaction rates. This is dictated  by their correct high temperature behaviour. 

The neutrino thermal masses, on the  other hand, can be neglected since these are suppressed by $\lambda^2$.

\subsection{Constraints on sterile neutrino dark matter}

\begin{figure}[!htb]
\begin{center}
\includegraphics[angle=-90,width=0.8\linewidth]{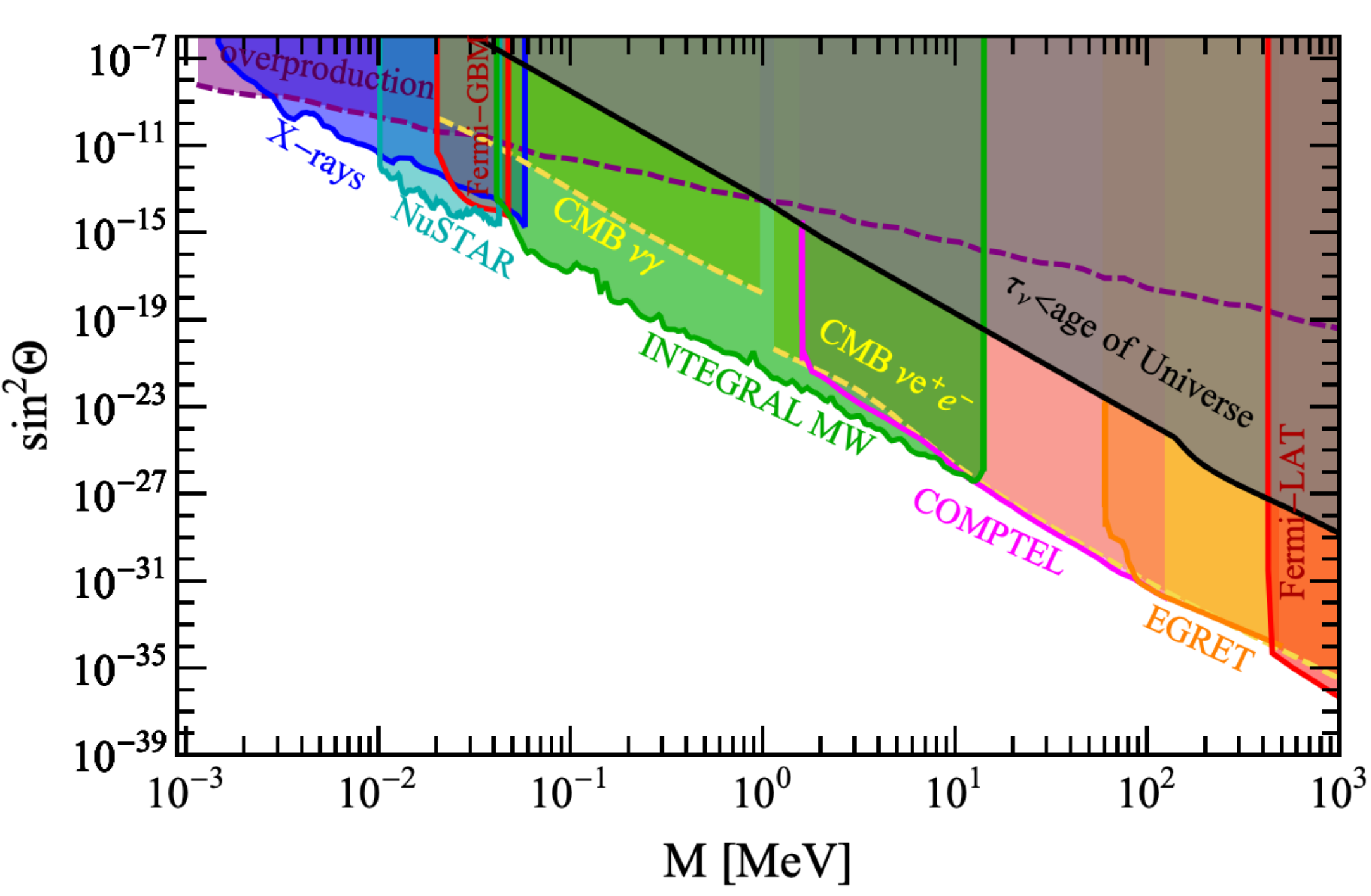}
\caption{ Neutrino dark matter constraints  on  the active--sterile mixing angle $\Theta$. The shaded areas are excluded by: X-ray data (dark blue), NuSTAR (dark cyan),  Fermi-GBM (red), INTEGRAL (green), COMPTEL (magenta), EGRET (orange), Fermi-LAT (red). In the dark grey region, the sterile neutrino lifetime is shorter than the age of the Universe. Above the purple dashed line the sterile neutrino is overabundant (assuming production only via mixing with active neutrinos).  The CMB constraints are given by the yellow dashed lines. }
\label{fig:nus_parspace}
\end{center}
\end{figure}

In Fig.~\ref{fig:nus_parspace}, we collect the most stringent limits   on the active--sterile mixing $\sin^2 \Theta$ as a function of the sterile neutrino mass $M$. We assume the abundance of sterile neutrinos to be equal to the dark matter density measured by Planck~\cite{Ade:2015xua,Aghanim:2018eyx}. 
Since our dark matter candidate decays into active neutrinos and other SM states, there are various strong constraints on this scenario.
The  most relevant $\nu$ decay modes are~\cite{PhysRevD.25.766,Gorbunov:2007ak,Essig:2013goa,Adhikari:2016bei}:

\begin{eqnarray}
\label{eq:sterilenudecays}
\Gamma_{\nu_a \gamma} &=& \frac{9 \alpha_{EM} G_f^2 M^5 }{256 \pi^4} \rm sin^2 \Theta\,,\\
\Gamma_{\nu_a e^+ e^-} &=&  \theta(M-2m_e)\frac{G_f^2 M^5}{96 \pi^3}  \rm sin^2 \Theta \frac{(1 + 4 sin^2 \theta_w + 8 sin^4 \theta_w)}{4}\,, \nonumber \\
\Gamma_{\nu_a \pi^0} &=&  \theta_{\rm }(M-m_{\pi^0})\frac{G_f^2 M^3 f_{\pi^0}^2}{32 \pi}  \rm sin^2 \Theta  \left(1-{m_{\pi^0}^2 \over M^2}\right)^2\,, \nonumber \\
\Gamma_{\nu_a\nu_a \bar{\nu}_a} &=& \frac{G_f^2 M^5}{96 \pi^3}  \rm sin^2 \Theta \nonumber\,,
\end{eqnarray}
where $\nu_a$ indicates an active neutrino. For a heavier $\nu$, further decay modes become relevant, e.g. those involving muons. Here we are assuming that the mixing  with the electron neutrino dominates. 

In the dark grey region, the sterile neutrino lifetime  is shorter than the age of the Universe. Sterile neutrinos are always produced in a thermal bath via the sterile--active mixing.
This leads to the ``overproduction'' constraint indicated by the dashed  purple line, above which the sterile neutrino abundance exceeds that of dark matter.

The sterile neutrino radiative decay is particularly relevant for X-ray and gamma-ray line searches. For sterile neutrino masses $M \lesssim 50$ keV, searches of decaying dark matter signals have been carried out using a wide range of X-ray telescopes like XMM-Newton~\cite{Boyarsky:2005us,Watson:2006qb}, Suzaku~\cite{Loewenstein:2008yi},  HEAO-1~\cite{Boyarsky:2005us},  INTEGRAL~\cite{Yuksel:2007xh,Boyarsky:2007ge}, Swift~\cite{Mirabal:2010an} and CHANDRA~\cite{RiemerSorensen:2009jp,Horiuchi:2013noa}. We collect most of them in the dark blue shaded area.\footnote{This bound takes into account the uncertainty in the dark matter density, as in~\cite{Boyarsky:2018tvu}.} Among them, the CHANDRA satellite provides the strongest limits~\cite{Perez:2016tcq}. Most recent bounds from the X-ray microcalorimeter NuSTAR~\cite{Roach:2019ctw}, looking at the Galactic Bulge, are displayed in dark cyan. The limits from searches for sterile neutrino decay lines using the Gamma-ray Burst Monitor onboard the Fermi Gamma-Ray Space Telescope (Fermi-GBM)~\cite{Ng:2015gfa} are shown in red. The green region is further constrained by INTEGRAL~\cite{Boyarsky:2007ge} searching for spectral lines from dark matter with a mass up to 14 MeV, decaying in the Milky Way halo.
Gamma-ray lines searches further constrain our sterile neutrino dark matter parameter space at higher masses: we show the bounds from COMPTEL~\cite{COMPTEL,Essig:2013goa} (magenta), EGRET~\cite{Strong:2004de,Essig:2013goa} (orange) and  Fermi Large Area Telescope (Fermi-LAT)~\cite{Ackermann:2015lka} (red).

Finally, measurements of the cosmic microwave background (CMB) allow us to constrain sterile neutrino decays leading to early energy injections
\cite{Adams:1998nr,Ichiki:2004vi,DeLopeAmigo:2009dc,Audren:2014bca,Poulin:2016anj}. The relevant decay modes are
$ \nu_a e^+ e^-$ and to $\nu_a \gamma$. Using the bounds on the corresponding decay rates from 
 Ref.~\cite{Slatyer:2016qyl} with appropriate photon flux rescaling, we obtain the CMB bounds shown by the yellow dashed lines.

We see that the resulting constraints on the mixing angle are very strong. For example, for $M $ close to 1 GeV, the bound on $\Theta$ is of order $10^{-18}$. Such small
values appear unnatural.   
  Within our simple model, the  angles are input parameters, while in various extensions their small values can be justified by {\it flavor--dependent} symmetry. Indeed, in addition to the lepton number $Z_4$, one may   impose a $Z_2$ symmetry
which acts on the lightest Majorana neutrino $\nu_{R_1}$:
 \begin{equation}
  \nu_{R_1} \rightarrow -    \nu_{R_1}  ~~\Rightarrow ~~ \Theta =0 \;.
    \end{equation}
This forbids the corresponding Yukawa couplings and sets $\Theta=0$. Assuming that this $Z_2$ is broken at some scale, the effective Yukawa couplings can be generated by higher dimensional operators. As a result, very small mixing angles can be generated. Since in the limit $\Theta \rightarrow 0$ the system becomes more symmetric, small mixing angles are natural according to the t'Hooft criterion \cite{tHooft:1979rat}. 

Since the lightest sterile neutrino effectively decouples from the Standard Model, 
the active neutrino masses are generated by the heavier $\nu_{R_{i}}$. The usual seesaw result for the active neutrino mass matrix $M_\nu$ still applies: $M_\nu = (M_D)^T M^{-1} M_D$, where $M_D$ is the Dirac mass matrix and $M$ is the diagonal Majorana mass matrix. Since we leave  the total number of sterile neutrinos arbitrary, no relevant, model independent, constraints are imposed by the low energy neutrino data
\cite{Asaka:2005an}.

 Longevity of the lightest sterile neutrino can be achieved at  small masses and/or small mixings. While most research efforts have focused on the first option, here we are considering the second possibility in more detail. We also see that, given the vast $(\Theta,M)$ parameter space, dark matter decay may be observed, e.g.  via monochromatic X- or gamma rays.

 \section{Relativistic rates for asymmetric reactions }
 \label{asym}

Neutrino dark matter can be produced through a number of reactions. These include both scattering and decay which take place in the relativistic regime, i.e. when the temperature exceeds the particle masses. Since there are bosons in the initial state,  relativistic Bose--Einstein enhancement can be very significant and the reaction rates must take it into account.
The relevant results for symmetric reactions, that is, involving particles with the same mass in the initial state, have been obtained in  \cite{Arcadi:2019oxh},\cite{Lebedev:2019ton}.
In our case, some reactions can be asymmetric, e.g. $H+S \rightarrow X$, and these results must be generalized to particles of different masses.

In this section, we generalize the  relativistic  reaction rates based on the Bose--Einstein statistics  \cite{Arcadi:2019oxh},\cite{Lebedev:2019ton} to processes involving particles with different masses.  
The $a \rightarrow b$ reaction rate
per  unit volume  is given by the general expression
\begin{equation}
\Gamma_{a\rightarrow b} = \int \left( \prod_{i\in a} {d^3 {\bf p}_i \over (2 \pi)^3 2E_{i}} f(p_i)\right)~
\left( \prod_{j\in b} {d^3 {\bf p}_j \over (2 \pi)^3 2E_{j}} (1+f(p_j))\right)
\vert {\cal M}_{a\rightarrow b} \vert^2 ~ (2\pi)^4 \delta^4(p_a-p_b) . 
\label{Gamma}
\end{equation}
Here ${\cal M}_{a\rightarrow b}$ is the  QFT transition amplitude,  in which we also absorb the {\it  initial and final} state symmetry factors,
and $f(p)$ is the momentum distribution function. 
For the freeze--in scenario, the density of the final state particles is small so that the enhancement factors
 $1+f(p_j)$ can be set to one. On the other hand, it is important to keep the full Bose--Einstein distribution functions  $f(p_i)$ for the initial state and their replacement by the 
 Maxwell--Boltzmann ones can lead to a rate  underestimate by orders of magnitude.

We are particularly interested in the $2\rightarrow 2$ reactions. 
The reaction rate can be expressed in terms of the cross-section,
\begin{equation}
\Gamma_{2 2} = (2\pi)^{-6} \int d^3 {\bf p_1} d^3 {\bf p_2} ~f(p_1) f(p_2) ~\sigma (p_1,p_2) v_{\rm M \o l}
\end{equation}
with 
\begin{equation}
v_{\rm M\o l}= {F \over E_1 E_2} \equiv { \sqrt{(p_1 \cdot p_2)^2 - m_1^2 m_2^2} \over E_1 E_2 } \;,
\end{equation}
\begin{equation}
f(p)= {1 \over \exp^{u\cdot p \over T} -1 } ~ ~, ~~u=(1,0,0,0)^T\;.
\end{equation}
The cross section is defined by
\begin{equation}
\sigma (p_1,p_2)= {1\over 4 F(p_1,p_2)} \int \vert {\cal M} \vert^2 (2\pi)^4 \delta^4 \left(p_1+p_2 -\sum_i k_i\right)
\prod_i {d^3 {\bf k}_i \over (2 \pi)^3 2E_{k_i}}  \;,
\label{sigma-def}
\end{equation}
where ${\cal M}$ is the QFT transition amplitude. 
Here we absorb  the symmetry factors for the {\it initial state} directly  into 
$\sigma(p_1,p_2)$.
 
The calculation is most easily performed in the center--of--mass (CM) frame, so let us convert the integral into that frame.
  The CM frame for each pair $p_1,p_2$ is the frame where $p_1+p_2$ has only zero spacial components. Let us consider how the integration measure transforms as we go to the CM frame.
The Lorentz invariant measure is
\begin{equation}
{d^3 {\bf p_1}\over 2E_1} {d^3 {\bf p_2}\over2 E_2} = d^4 p_1 d^4 p_2~ \delta (p_1^2-m_1^2) \delta (p_2^2-m_2^2)\;.
\end{equation}
Introduce 
\begin{equation}
p={p_1+p_2\over 2} ~,~k={p_1-p_2\over 2} ~,
\end{equation}
such that 
\begin{equation}
d^4 p_1 d^4 p_2~ \delta (p_1^2-m_1^2) \delta (p_2^2-m_2^2)= 2^4 d^4p ~d^4k ~\delta((p+k)^2-m_1^2) \delta((p-k)^2-m_2^2) \;.
\end{equation}
Any time-like vector $p$ can be Lorentz-transformed to the form 
\begin{equation}
p=\Lambda(p)~\left( 
\begin{matrix}
&E& \\
&0&\\
&0&\\
&0&
\end{matrix}
\right),
\end{equation}
with the explicit parametrization in terms of rapidity $\eta$ and angular coordinates $\theta,\phi$ being
\begin{eqnarray}
&& p^0= E \cosh \eta, \nonumber\\ 
&& p^1=E \sinh \eta \sin \theta \sin \phi ,\nonumber\\
&& p^2=E \sinh \eta \sin \theta \cos \phi ,\nonumber\\
&& p^3=E \sinh \eta \cos \theta  .\nonumber\\
\end{eqnarray}
In other words,   in the convention  $p=(p^0, p^3, p^2, p^1)^T$, we have 
\begin{eqnarray}
&&\Lambda(p)=\left( 
\begin{matrix}
&1 & 0& 0& 0& \\
&0 & 1& 0& 0&  \\
&0 & 0& \cos\phi & -\sin\phi&  \\
&0& 0& \sin\phi & \cos\phi &
\end{matrix}
\right)
\left( 
\begin{matrix}
&1 & 0& 0& 0& \\
&0 & \cos\theta & -\sin\theta & 0&  \\
&0 & \sin\theta & \cos\theta & 0&  \\
&0& 0& 0 & 1 &
\end{matrix}
\right)
\left( 
\begin{matrix}
&\cosh \eta & \sinh \eta& 0& 0& \\
&\sinh \eta & \cosh \eta & 0 & 0&  \\
&0 & 0 & 1 & 0&  \\
&0& 0& 0 & 1 &
\end{matrix}
\right),\nonumber \\
&& \Lambda(p)^{-1}= 
\left( 
\begin{matrix}
&\cosh \eta & -\sinh \eta& 0& 0& \\
&-\sinh \eta & \cosh \eta & 0 & 0&  \\
&0 & 0 & 1 & 0&  \\
&0& 0& 0 & 1 &
\end{matrix}
\right) 
\left( 
\begin{matrix}
&1 & 0& 0& 0& \\
&0 & \cos\theta & \sin\theta & 0&  \\
&0 & -\sin\theta & \cos\theta & 0&  \\
&0& 0& 0 & 1 &
\end{matrix}
\right)
\left( 
\begin{matrix}
&1 & 0& 0& 0& \\
&0 & 1& 0& 0&  \\
&0 & 0& \cos\phi & \sin\phi&  \\
&0& 0& -\sin\phi & \cos\phi &
\end{matrix}
\right). \nonumber
\end{eqnarray}

The $p$-vector in the form $(E,0,0,0)^T$ corresponds to the CM frame and $E>0$ is half the CM energy. 
The $p$-integration measure   becomes
\begin{equation}
d^4p= \sinh^2 \eta  E^3dE ~d\eta ~d\Omega_p \;,
\end{equation}
where $\Omega_p$ is the solid angle in $p$-space.
Now apply the same Lorentz transformation $\Lambda(p)$ to the vector $k$,
\begin{eqnarray}
  k= \Lambda(p)~k' &\xrightarrow{\text{ drop~the~prime}}& k ,\nonumber\\
 d^4k=d^4k' &\xrightarrow{\text{ drop~the~prime}}&  d^4k \equiv dk_0 \vert {\bf k}\vert^2 d \vert {\bf k}\vert  d\Omega_k \;,
\end{eqnarray}
where we have used the fact that $\Lambda(p)$ is a constant Lorentz transform with respect to the variable $k$
so that the measure remains invariant. We drop the prime for convenience, remembering that $k$ now is in
{\it the CM frame}. $\Omega_k$ denotes the corresponding solid angle in that frame.

Let us now integrate the delta functions. We can explicitly integrate over $k_0$ and $\vert {\bf k}\vert$.
In the CM frame, the delta functions   become
\begin{equation}
\delta(E^2 +2Ek_0+k_0^2-{\bf k}^2-m_1^2)~\delta(E^2 -2Ek_0+k_0^2-{\bf k}^2-m_2^2) \;.
\end{equation}
This enforces
\begin{eqnarray}
&& k_0 = {m_1^2-m_2^2 \over 4E} \;,  \nonumber \\
&& {\bf k}^2 = E^2 - {m_1^2+m_2^2 \over 2} + {(m_1^2-m_2^2)^2\over 16E^2} \;.
\label{k0}
\end{eqnarray}
We then have 
\begin{eqnarray}
&& \int dk_0 d \vert {\bf k}\vert \vert {\bf k}\vert^2 ~\delta(E^2 +2Ek_0+k_0^2-{\bf k}^2-m_1^2)~\delta(E^2 -2Ek_0+k_0^2-{\bf k}^2-m_2^2) = 
  {\vert {\bf k} \vert \over 8E}  \;, \nonumber
\end{eqnarray}
 which allows us to rewrite the integration measure as
\begin{equation}
\int {d^3 {\bf p_1}\over 2E_1} {d^3 {\bf p_2}\over2 E_2} ...=
{1\over 2} \int_{m_1+m_2\over 2}^\infty  dE  ~E  ~\sqrt{(4E^2- m_1^2 -m_2^2)^2 -4m_1^2 m_2^2  }   \int_0^\infty  d\eta  ~\sinh^2 \eta  ~ \int d\Omega_p ~d\Omega_k ... \;,
\end{equation}
where in the integrand one must set  $k_0$ and $\vert {\bf k} \vert $  to their values given by Eq.\;(\ref{k0}). 
Note that $E$ is half the CM energy.

Since the cross section in the CM frame is a function of $E$ only, 
the angular dependence comes entirely from the distribution functions. We have 
\begin{eqnarray}
&&u\cdot p_1= (\Lambda^{-1}u) \cdot (p+k)= (E+k_0)\cosh\eta + \vert {\bf k}\vert \sinh\eta  ~\cos\theta_k \;,\nonumber\\
&&u\cdot p_2= (\Lambda^{-1}u) \cdot (p-k)= (E-k_0)\cosh\eta - \vert {\bf k}\vert  \sinh\eta  ~\cos\theta_k \;.
\end{eqnarray}
Here we have used  $k^3= \vert {\bf k}\vert \cos \theta_k$. 

Integration over $\Omega_p$ gives $4\pi$ and the integral over $\phi_k$ is $2\pi$. Let us now integrate over $\theta_k$.
The integral can be reduced to 
\begin{eqnarray}
 I_\theta=  \int_{-1}^1 dx ~{1\over e^{a+bx}-1} {1\over e^{c-bx}-1}&=& 
   {1\over b(e^{a+c}-1)} \ln {  \left[ { \sinh {a+b\over 2}  \over  \sinh {a-b\over 2}} ~  {\sinh {c+b\over 2}  \over  \sinh {c-b\over 2}}   \right] }
\label{Itheta}
\end{eqnarray}
for $a,c >b$. Here $a={(E+k_0) \cosh \eta \over T}$,  $c={(E-k_0) \cosh \eta \over T}$ and 
$b= {\vert {\bf k} \vert \sinh\eta \over T}$. (This result can most easily be obtained by the change of variables to $y = e^{bx}$.)

We thus get 
\begin{eqnarray}
&&\Gamma_{22} = (2\pi)^{-6} \int {d^3 {\bf p_1}\over 2E_1} {d^3 {\bf p_2}\over2 E_2} ~f(p_1) f(p_2) ~4F(p_1,p_2) \; \sigma(p_1,p_2) \nonumber   \\
&& = {T\over 4 \pi^4 }  \int_{m_1+m_2\over 2}^\infty  dE  ~E^2   \int_0^\infty  d\eta {    \sinh \eta \over e^{{2E\over T} \cosh\eta }-1}~
\ln  \left[ {     \sinh    {(E+k_0) \cosh\eta + \vert {\bf k}\vert  \sinh\eta \over 2T}    \over
\sinh   {(E+k_0) \cosh\eta - \vert {\bf k}\vert   \sinh\eta \over 2T}   } ~  {     \sinh    {(E-k_0) \cosh\eta + \vert {\bf k}\vert  \sinh\eta \over 2T}    \over
\sinh   {(E-k_0) \cosh\eta - \vert {\bf k}\vert   \sinh\eta \over 2T}   } \right]  \nonumber \\
&& \times 4F \sigma^{\rm CM} (E) \;,
\label{22rate}
 \end{eqnarray} 
  with $k_0$ and $\vert {\bf k}\vert $ given by (\ref{k0}).
   This expression  reduces to that  of    \cite{Arcadi:2019oxh},\cite{Lebedev:2019ton} for equal masses, $m_1=m_2$. 
  
  It is important to note that the masses here must include thermal corrections (\ref{ther-mass}). This is necessary for the correct high temperature behaviour:
  \begin{equation}
  \Gamma_{22} \propto T^4 \ln {T\over m} \rightarrow {\rm const}\;T^4 
  \end{equation}
 only when a thermal correction to $m$ has been included. The latter also regularizes the infrared divergence in the massless limit.
  
\section{Thermalization constraints}
\label{thermalization}

 In this work, we focus on freeze--in production of sterile neutrinos. Freeze--in calculations are reliable only if the produced particles do not thermalize. 
 This requires the coupling between the thermal bath and the frozen--in particles as well as self--interaction of the latter to be sufficiently small. In this section, we delineate parameter space consistent with these conditions. We use relativistic formulas for the reaction rates, taking into account quantum statistics for the initial state.
 
 Let us consider the regime 
 where  $H$ and $S$ develop VEVs $v$ and $u$, respectively.  We can parametrize them in the unitary gauge as
  \begin{eqnarray}
  && H \rightarrow {1\over \sqrt{2}} (h+v) \;, \nonumber\\
  && S \rightarrow s + u \;.
  \end{eqnarray}
  In terms of the {\it 4--component Majorana neutrino $\nu$}, the relevant to our calculation terms in the Lagrangian are 
  \begin{eqnarray}
    -\Delta {\cal L}_1& =& {1\over 2} \lambda\; s \; \bar \nu \nu +  {1\over 2} M \; \bar \nu \nu \;, \label{interactions} \\
      -\Delta {\cal L}_2 & =&
     {1\over 2} m_h^2 h^2 + {1\over 2} m_s^2 s^2 + 
    {v\over 2} \lambda_{hs} hs^2 + {u\over 2} \lambda_{hs} sh^2 + u \lambda_s s^3 
    + {1\over 4} \lambda_{hs} h^2 s^2 + {1\over 4 } \lambda_s s^4\;, \nonumber
  \end{eqnarray}
  where $M=\lambda u$ and we have neglected the scalar mixing. The Majorana notation has the advantage that the $\nu \nu$ final state includes all combinations of 2--component neutrinos and anti--neutrinos.

 \subsection{Sterile neutrino thermalization }

We show in Sec.\;\ref{secI} that the main production channel for sterile neutrinos is  the decay $s \rightarrow \nu\nu$. Here we assume that $s$ is in thermal equilibrium and
$m_s \gg M$. For  a sufficiently large $\lambda$, the decay is efficient and the neutrino number density $n_\nu (T)$ approaches its equilibrium value at a given temperature,
$n_\nu^{\rm eq} (T)$.
In this case, the reverse process $\nu\nu \rightarrow s$ becomes important and the neutrinos tend to equilibrate with the thermal bath of $s$. Thus, we use the 
non--thermalization criterion
\begin{equation}
n_\nu (T) < n_\nu^{\rm eq} (T)
\label{non-thermalization}
\end{equation}
for any $T$ down to temperatures around  $M/3$. (At lower $T$, $n_\nu^{\rm eq} (T)$ is exponentially suppressed.)

The number density $n_\nu$ is calculated via the Boltzmann equation 
 \begin{equation}
 \dot{n}_\nu +3n_\nu H = 2  \Gamma_{12} (s\rightarrow \nu\nu)   \;,
\end{equation}
where $H$ is the Hubble rate, 
  \begin{equation}
 H= \sqrt{ \pi^2 g_* \over 90} \; {T^2 \over M_{\rm Pl}} \;,
\end{equation}
 with $g_*$ being the number of active SM degrees of freedom.
$\Gamma_{12} (s\rightarrow \nu\nu)$ is the     reaction rate per unit volume (see Section \ref{secI} for an explicit expression). It  is calculated with the Bose--Einstein distribution for $s$,
while neglecting the final state Pauli blocking due to the low density of $\nu$, as is usual in freeze--in computations. 
Since  the issue of {\it bona fide} thermalization is quite complicated in any case,  this approximation is adequate for our purposes.

  \begin{figure}[h!]
\begin{center}
 \includegraphics[scale=0.45]{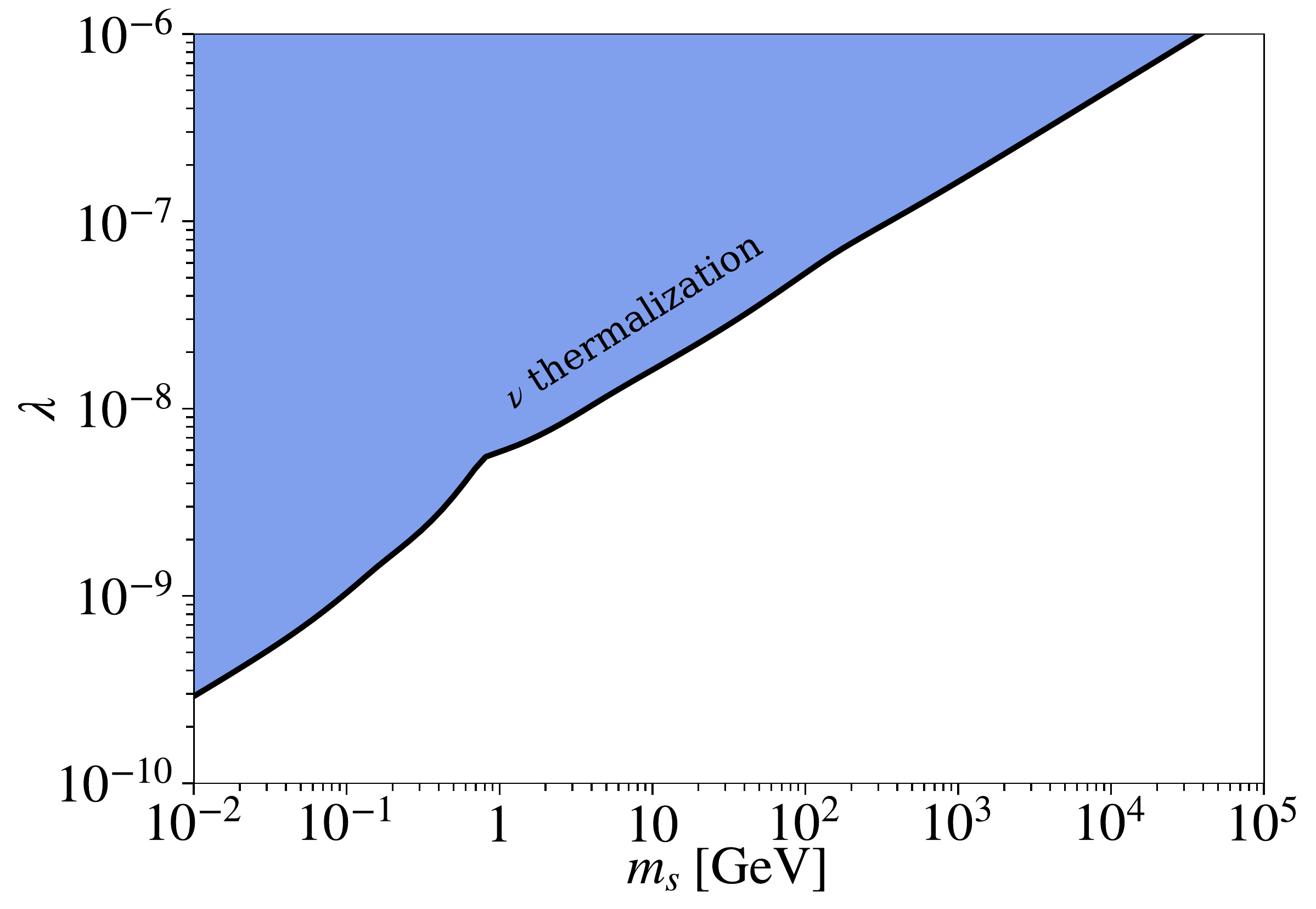}
\end{center}
\caption{   Sterile neutrino non--thermalization bound. The neutrinos are produced via 
 $s\rightarrow \nu\nu$ with $s$ in thermal equilibrium and 
  $m_s \gg M$.  In the shaded region, $n_\nu \gtrsim n_\nu^{\rm eq}$
such that the reverse process $\nu\nu \rightarrow s$ is important.}
\label{nu-therm}
\end{figure}
 
The solution to the Boltzmann equation for fixed  $\lambda, m_s$  and zero initial $n_\nu $ is then compared to the equilibrium neutrino density at  a given $T$. If inequality
(\ref{non-thermalization}) is satisfied for any $T$ above $M/3$, the freeze--in approximation is adequate. 
Our numerical results for $m_s \gg M$ are presented in Fig.\;\ref{nu-therm}. In the shaded region, the neutrino density equals or exceeds its equilibrium value. 
The kink at roughly 1 GeV appears due to the significant change in $g_*$ at the QCD phase transition.
We see that only quite small couplings, e.g. below $10^{-8}$ at $m_s \sim 1 $ GeV, are consistent with the freeze--in approximation.
 The bound  can be approximated by
 \begin{equation}
 \lambda <  5 \times 10^{-9} \;  \sqrt{m_s \over   {\rm GeV}} \;.
 \end{equation}
 The qualitative behaviour of $\lambda(m_s)$ can be understood from $\lambda^2$-dependence of the rate and the fact that the main contribution to $n_\nu$ comes from
 temperatures of order $m_s$. We discuss this in more detail in Sec.\;\ref{sec:abundance}.
 
 In the vicinity of the shaded region, the neutrino density is significant such that the final state quantum statistics (Pauli blocking) can have a tangible impact on the reaction rate. This effect would reduce the rate, hence our bound is somewhat more restrictive than the true one  and can be viewed as conservative.

Let us note that other possible ``thermalization'' conditions appear in the literature. For example, one can compare the neutrino production rate to the Universe expansion rate.
If the former dominates, one expects the neutrino sector to be quickly populated. 
In our case, this corresponds to $n_s^{-1} \Gamma_{12} (s\rightarrow \nu\nu) \gtrsim 3H$. While such a condition  often leads to similar results, there are notable exceptions.
In particular, the above inequality  is always satisfied at low enough temperatures regardless of the coupling. This, however, does not mean that the neutrino sector gets populated. Indeed, when $s$ is non--relativistic, 
$n_s^{-1} \Gamma_{12} (s\rightarrow \nu\nu)$ is approximately constant, while $H$ decreases as $T^2$. As a result, all the $s$--quanta  available at the corresponding temperature get converted into $\nu$ pairs. Yet, since for relativistic neutrinos $n_s^{\rm eq} (T) \ll n_\nu^{\rm eq}  (T)$, the neutrino density increase is negligible and $\nu$'s do not thermalize.
Another exception is the situation in which the production is intense but short in duration, e.g. around a phase transition. In this case, the accumulated density can still be small.

 \subsection{Thermalization of $s$}
 
In this work, we assume that the dominant source of $s$--quanta is the Higgs thermal bath. It is  important to understand under what circumstances the processes
$h \leftrightarrow ss$, $hh \leftrightarrow ss$ and $hh \leftrightarrow s$ lead to thermalization of $s$.  As in the previous section, we use the criterion
\begin{equation}
n_s (T) < n_s^{\rm eq} (T)
\label{non-thermalization}
\end{equation}
for any $T \gsim m_s$ to ensure non--thermalization of $s$.\footnote{In practice, we check this condition down to temperatures $T \sim m_s/3$, where $s$ becomes non--relativistic.} The number density $n_s$ is calculated numerically via the Boltzmann equation
\begin{equation}
 \dot{n}_s+3n_s H = \sum_i a_i \Gamma_i   \;,
\end{equation}
 where $\Gamma_i$ are the reaction rates $\Gamma_{12}(h \rightarrow ss)$, $\Gamma_{22}(hh \rightarrow ss)$, $\Gamma_{21}(hh \rightarrow s)$ and
 $a_i$ take into account the number of $s$--particles in the final state as well as the number of Higgs d.o.f. The explicit expressions for the rates are given 
 in Sec.\;\ref{freeze-in-s}. \\ \ \\

  \begin{figure}[h!]
\begin{center}
 \includegraphics[scale=0.45]{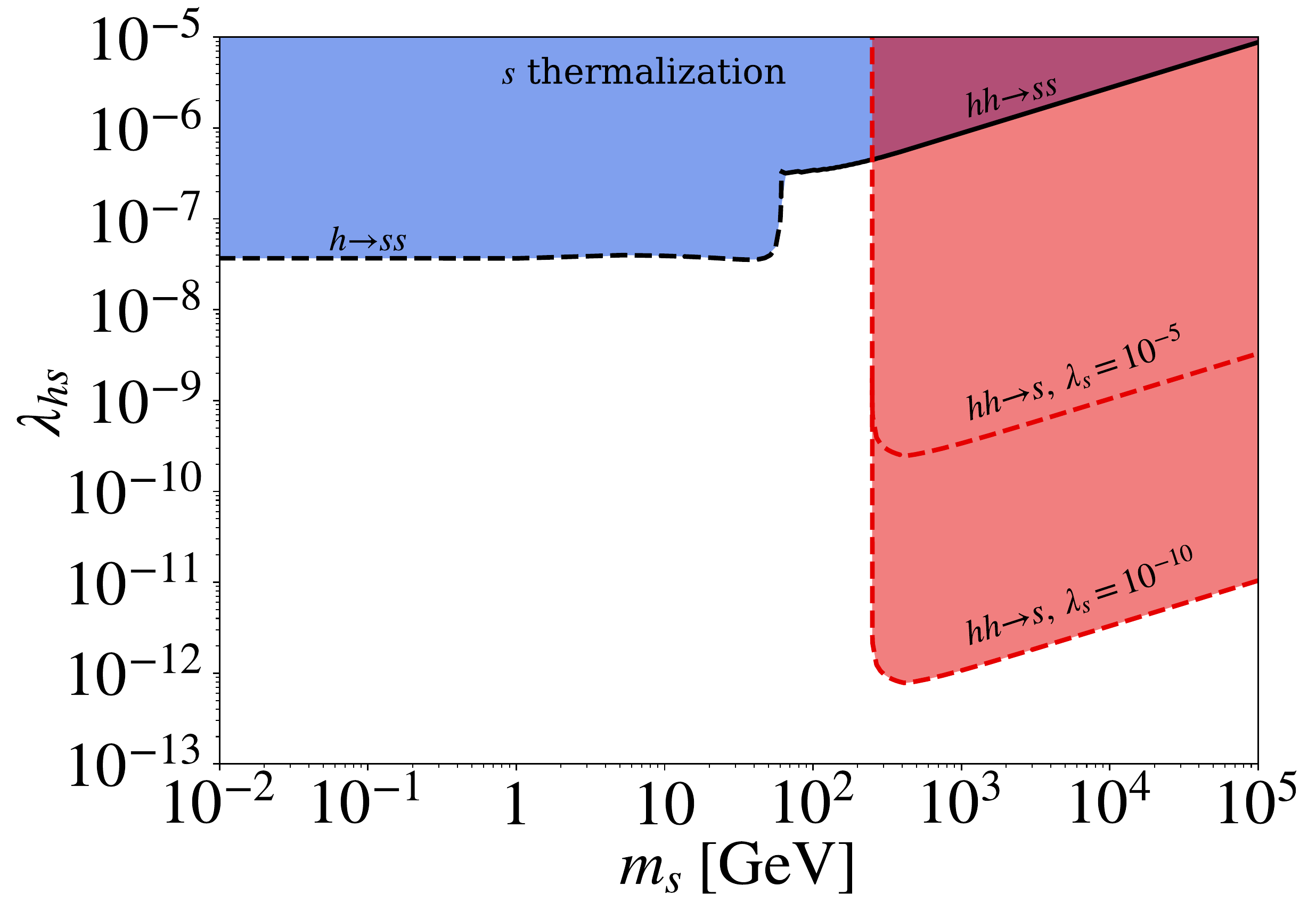}
\end{center}
\caption{ Upper bounds on $\lambda_{hs}$ from non--thermalization of $s$. 
 In the shaded regions, $n_s (T) \geq n_s^{\rm eq} (T)$ due to the $h \rightarrow ss$, $hh \rightarrow ss$ and $hh \rightarrow s$ processes.
 Here the scalar mixing and the electroweak (EW) transition effects have been neglected. The fusion mode $hh \rightarrow s$ is sensitive to $\lambda_s$, for which two benchmark values 
 $10^{-5}$ and $10^{-10}$ have been chosen. }
\label{lambda-hs-therm}
\end{figure}

  The resulting bounds on $\lambda_{hs}$ are shown in Fig.\;\ref{lambda-hs-therm}. For a light $s$, the decay mode $h\rightarrow ss$ dominates, while for a heavy scalar 
 the production is typically dominated by the fusion mode $hh \rightarrow s$. The latter is sensitive to the $s$--VEV $u=m_s/\sqrt{2 \lambda_s}$, so additional input such as the coupling $\lambda_s$
 is required. This VEV grows very large at small $\lambda_s$ which results in a large reaction rate. Note that  in the vicinity of the shaded region, the final state Bose--Einstein enhancement factor
 can be non--negligible, so our procedure overestimates somewhat the upper bound on the coupling.

 The bound on $\lambda_{hs} $ at $m_s \ll m_h$ is independent of $m_s$,
 \begin{equation}
\lambda_{hs}(h\rightarrow ss) < 4 \times 10^{-8}\;.
\end{equation}
This is because $\Gamma_{12}(h \rightarrow ss)$ is independent of $m_s$ in this regime and the production stops around $T \sim m_h/5$. At larger $m_s$,
the scattering reaction $hh \rightarrow ss$ becomes significant.
The rate $\Gamma_{22}(hh \rightarrow ss)$ scales as $T^4$ in the relativistic regime and the resulting $n_s(T) \propto T^2$. The yield is dominated by low temperatures consistent with  the relativistic scaling, that is, $T \sim m_s$. We thus obtain
 \begin{equation}
\lambda_{hs}(hh \rightarrow ss) < 6 \times 10^{-8}\; \sqrt{ m_s\over {\rm GeV}}\;.
\end{equation}
 The fusion channel $hh \rightarrow s$ is more complicated. For $m_s \gg 2 m_h$, it becomes active at temperatures below $T \sim m_s$, that is, when the Higgses still have enough energy to produce $s$ and the Higgs thermal mass is {\it not too large} for the process to be kinematically allowed. The fusion becomes inefficient below $T \sim m_s/6$.
 In this regime, the reaction rate does not follow any simple scaling law and numerically we obtain
  \begin{equation}
\lambda_{hs}(hh \rightarrow s) < 6 \times 10^{-9}\; \sqrt{ \lambda_s m_s\over {\rm GeV}}\;.
\end{equation}
  The appearance of $\lambda_s$ can be understood from the reaction rate scaling as $\lambda_{hs}^2/\lambda_s$ for a fixed $m_s$.

 \subsubsection{Self--thermalization due to $\lambda_s$}
 
 Even if $\lambda_{hs}$ is small, the $s$--sector can thermalize due to self--interaction $\lambda_s s^4$. This happens when the number changing processes $ss \leftrightarrow ssss$ 
 become efficient and the density $n_s$ starts being  sensitive to $\lambda_s$. The specifics of self--thermalization are computationally involved. In the symmetric phase $u=0$
 at large $s$--densities close to equilibrium, the   (necessary) thermalization condition  on $\lambda_s$ has been derived in \cite{Arcadi:2019oxh}. Here we are interested in the broken phase $u\not=0$ at low $s$--densities and thus have to  resort to simple estimates. We assume that the initial $n_s$ is created via the Higgs thermal bath and study which values of $\lambda_s$ do not affect its evolution.

   \begin{figure}[h!]
\begin{center}
 \includegraphics[scale=0.45]{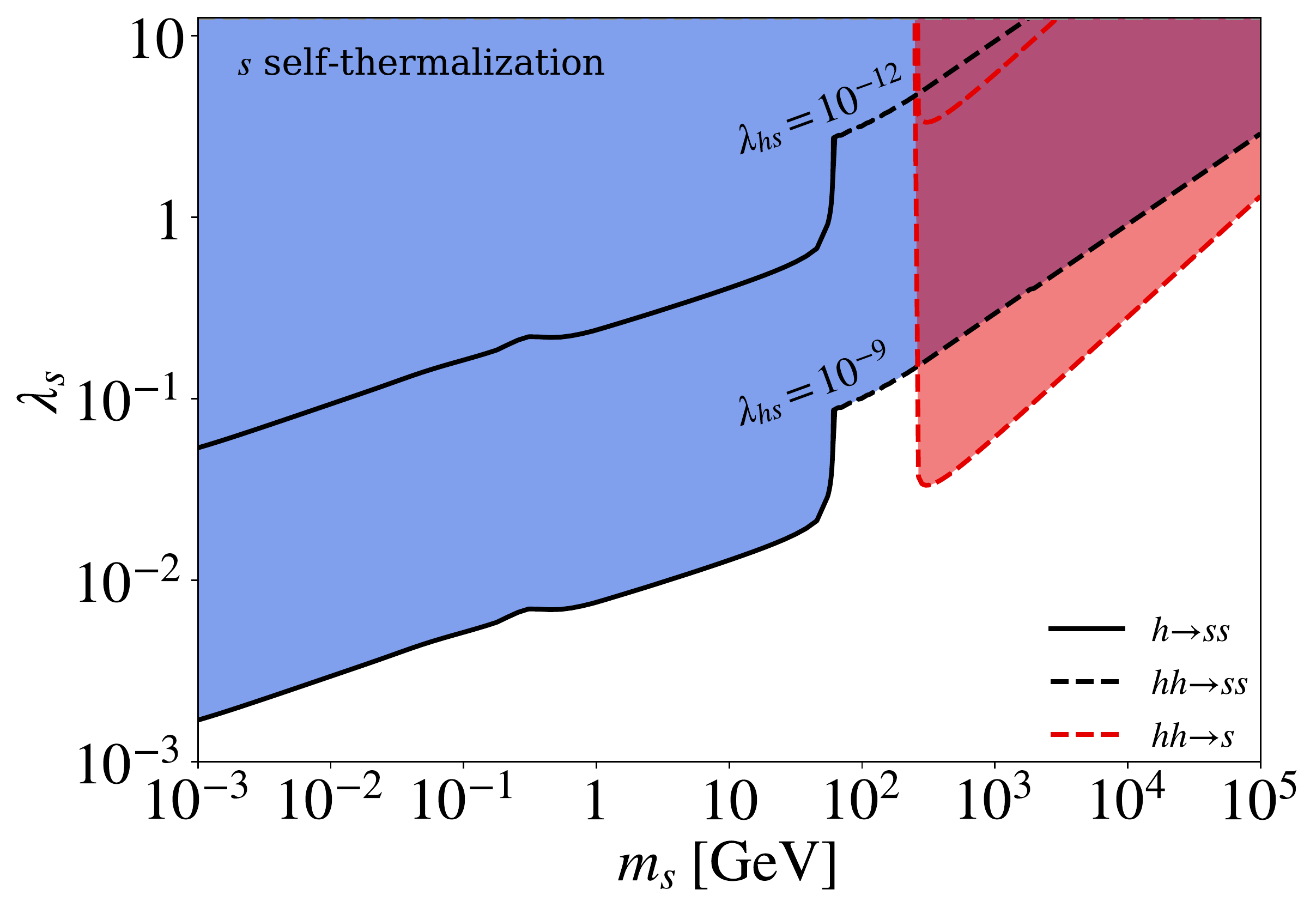}
\end{center}
\caption{ Estimates of the upper bound on $\lambda_{s}$ from non--thermalization of $s$. 
 In the shaded regions, the $ss \rightarrow ssss$ process is efficient. The bound depends on $n_s$ and thus is sensitive to $\lambda_{hs}$ and the 
 $s$--production mode. The displayed constraints correspond to two benchmark values of $\lambda_{hs}$: $10^{-12}$ and $10^{-9}$. 
 Here the scalar mixing and the EW transition effects have been neglected.  }
\label{lambda-s-therm}
\end{figure}
 
 The $2\rightarrow 4 $ reaction rate at low $s$--density can be written as
 \begin{equation}
 \Gamma_{24}= n_s^2 \langle \sigma_{24} v_{\rm rel}\rangle \;,
\end{equation}
 where $\sigma_{24}$ is the corresponding QFT cross section and $v_{\rm rel}$ is the relative (M\o ller) velocity. We are interested mostly in the relativistic regime, in which case 
 $\sigma_{24}(\hat s) \sim 10^{-4 } \lambda_s^4 \ln^2 (\hat s/2m_s^2) /\hat s$, where $\hat s \gg 4m_s^2$ is the Mandelstam variable. This result can be verified with CalcHEP \cite{Belyaev:2012qa}.
 For fixed  $\lambda_{hs}$ and $\lambda_s$,  
the density $n_s(T) \ll n_s^{\rm eq}(T) $ is calculated via the Boltzmann equation in the previous subsection.

Although the momentum distribution of $s$ is non--thermal, the characteristic  energy of the $s$--quanta can be approximated by $T$. This is because $n_s(T)$ is dominated by the late  time production in the Higgs thermal bath at temperature $T$.  In the relativistic regime, we may take $\hat s \sim 4T^2$ to calculate the average cross section and $v_{\rm rel} \sim 2$.
The number changing interactions are efficient if $2 \Gamma_{24} > 3 n_s H$, so to ensure non--thermalization we require 
 \begin{equation}
{  n_s \langle \sigma_{24} v_{\rm rel} \rangle  \over H} \Biggr\vert_{T\sim m_s}  \lesssim 1\;,
\label{s-self-therm}
\end{equation}
where we have taken into account the fact that this ratio is maximized at the lowest temperature consistent with the relativistic scaling.

 Our numerical results are shown in Fig.\;\ref{lambda-s-therm}.  Equation (\ref{s-self-therm}) makes it clear that the bound on $\lambda_s$ increases with $m_s$.  
Other qualitative features can be understood from the discussion in the previous subsection. We see that the upper bounds on $\lambda_s$ are significantly above those in
\cite{Arcadi:2019oxh} (cf. Fig.\;2). This is expected since the number density in our case is significantly below its equilibrium value.

 Let us emphasise that the above bounds  have been obtained under a number of simplifying assumptions. First of all, we have neglected the small scalar mixing, which is not expected to affect the results significantly. 
 We have also assumed that the density of produced particles is low enough such that the final state quantum statistics is unimportant.
 Finally, we have ignored EW phase transition effects. These can have a non--trivial impact on the bounds. In particular, as we discuss in 
 Sec.\;\ref{freeze-in-s}, the $hh\rightarrow s$  mode can be active even at light $m_s$ due to the Higgs mass reduction close to the critical temperature. In this sense, the presented constraints can be viewed as conservative.

\section{Sterile neutrino production I: thermalized $s$}
\label{secI}

\subsection{Reaction rates}

\begin{figure}[h!]
\begin{center}
 \includegraphics[scale=0.85]{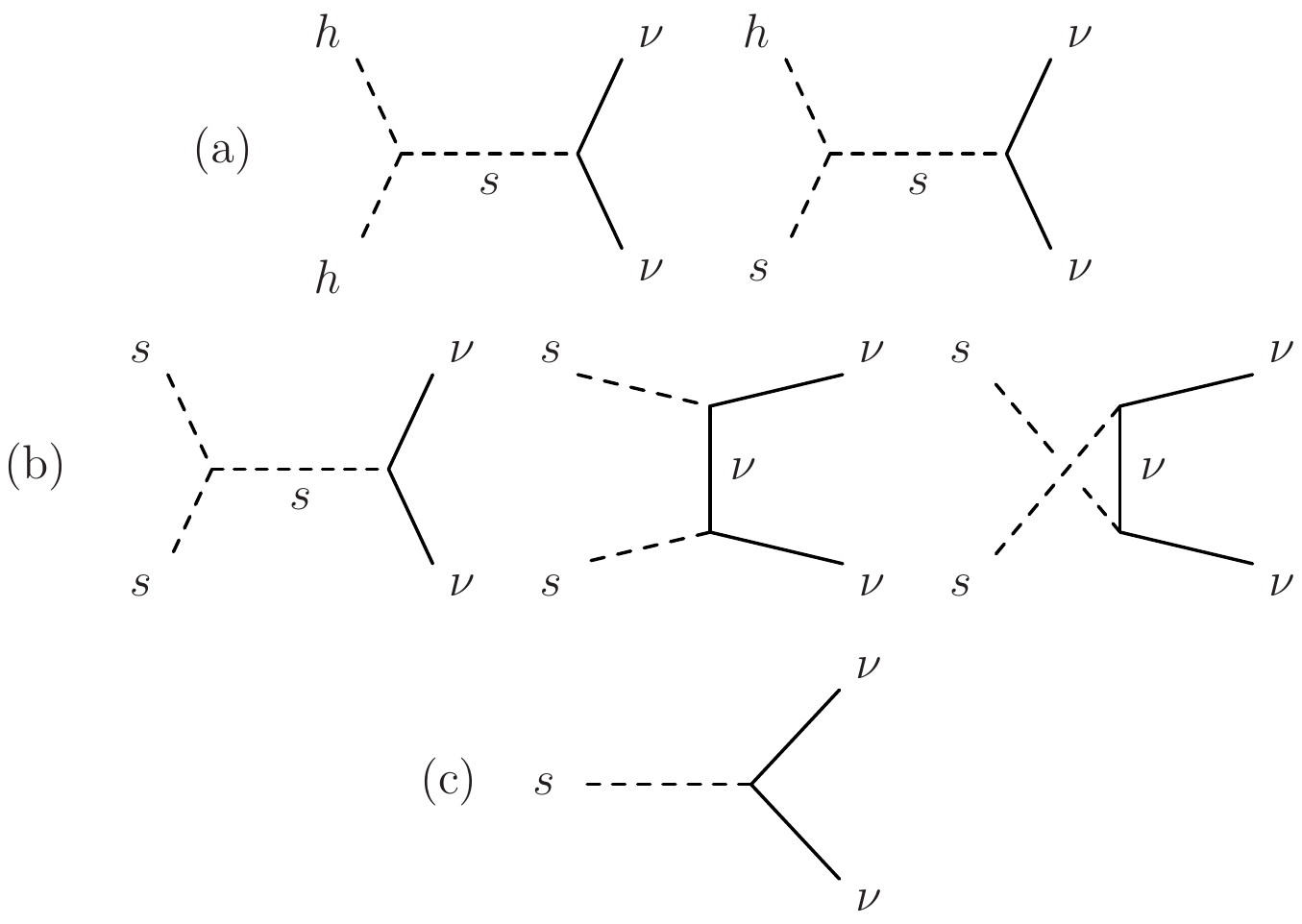}
\end{center}
\caption{   Neutrino dark matter production in a thermal bath: (a) $hh$ and $hs$ annihilation; (b) $ss$ annihilation; (c) $s$ decay.
(An analogous Higgs mode $h\rightarrow \nu\nu$ not shown.)}
\label{diagrams}
\end{figure}

 In the thermal bath of $h$ and $s$, there are a few channels for $\nu$ production, see Fig.\;\ref{diagrams}.
 The reactions $s\rightarrow \nu\nu$ and $ss\rightarrow \nu \nu$ take place at both high and low temperatures,
 while $hh \rightarrow \nu\nu$ and $hs\rightarrow \nu\nu$ require the presence of scalar trilinear vertices which only appear 
 below the corresponding critical temperatures.

 The relevant interactions are given by Eq.\;\ref{interactions}.  Note that the field VEVs and the degrees of freedom depend on the temperature.
   At high temperatures, the VEVs vanish, $u,v=0$,  and the single Higgs d.o.f. is replaced by 4 massive Higgs scalars $h_i$.
   In this work, we neglect the gauge boson contributions suppressed by an extra power of the gauge coupling (see e.g. \cite{Heeba:2018wtf}).
  We also neglect the scalar mixing $\theta \ll 1$ apart from the reaction $h \rightarrow \nu\nu$, which is absent at leading order in $\theta$.
  
  In what follows, we neglect the SM--like Yukawa coupling of the lightest sterile neutrino. As mentioned before, its tiny value can be justified by 
  the neutrino parity.
  
  Below we summarize our results for the reaction cross sections which are to be inserted into Eq.\;\ref{22rate} or its equal--mass analog.
  The masses that appear in the rates are meant to be the thermally corrected masses.

  \subsubsection{$hh \rightarrow \nu\nu$}
  
The calculation is easiest performed in the CM frame.
  The amplitude for the $\nu\nu$ final state is 
  \begin{equation}
  \vert {\cal M}\vert  =  \left\vert  {  u \lambda_{hs} \lambda \over \hat s-m_s^2} \; \bar u (p) v(p^\prime) \right\vert  \;.
  \end{equation}
  Here the combinatorial factor $2\times 2$ coming from two identical particles in the initial and final states is included; 
  $ {\hat s}=4E^2$ is the Mandelstam variable. The neutrino 4--momenta are denoted by $p$, $p^\prime$ and 
  $u,v$ are   4--spinors.

  The spin sum and phase space integration yield 
  \begin{equation}
   4F  \sigma^{\rm CM} (hh\rightarrow  \nu\nu) =   {\lambda^2 \lambda_{hs}^2 u^2 \over 16\pi  } \;
   {     (\hat s-4M^2)^{3/2}  \over \sqrt{\hat s} (\hat s -m_s^2)^2   } \;. 
     \end{equation}
 where in our convention we include $both$ the initial and final state phase space symmetry factors of $1/2$ in the cross section.
  
  \subsubsection{$hs\rightarrow  \nu\nu $}

  The corresponding amplitude is
  \begin{equation}
 \vert  {\cal M} \vert = \left\vert  {v \lambda_{hs} \lambda \over \hat s-m_s^2} \; \bar u (p) v(p^\prime) \right\vert \;.
  \end{equation}
  The resulting cross section is 
  \begin{equation}
   4F  \sigma^{\rm CM} (hs\rightarrow  \nu  \nu) =   {\lambda^2 \lambda_{hs}^2 v^2 \over 8\pi  } \;
   {   (\hat s-4M^2)^{3/2}   \over \sqrt{\hat s}(\hat s -m_s^2)^2   } \;. 
     \end{equation}
    As before, $\hat s\equiv 4E^2$, although $h$ and $s$  have different energies in the CM frame.
    
  \subsubsection{$ss\rightarrow \nu \nu$  }
  
  The process $ss\rightarrow \nu \nu$ can proceed both through the $s$-channel and the $t,u$-channels at 2d order in $\lambda$.
  The amplitude is
  \begin{eqnarray}
  \vert {\cal M}\vert  &=& \Biggl\vert  \;      \; {6u \lambda_s\lambda \over \hat s-m_s^2} \; \bar u (p) v(p^\prime) +
  \lambda^2 \;\bar u (p) \; {   \skew7\not{p} -\skew7\not{p_1} +M  \over \hat t-M^2 } v(p^\prime)  + \lambda^2 \;\bar u (p) \; {   \skew7\not{p} -\skew7\not{p_2} +M  \over \hat u-M^2 } v(p^\prime) 
    \Biggr\vert , \;\;\;\;\;
  \end{eqnarray}
  where $\hat t,\hat u$ are the Mandelstam variables. The 4--momenta of the initial state particles  are denoted by $p_1$ and $p_2$.
  
  The resulting cross section is

{\small
\begin{align}
 \sigma_{ss\to\nu\nu}^\mathrm{CM}=&
 \frac{\lambda^2}{16\pi \hat s}\frac{\sqrt{\hat s-4M^2}}{\sqrt{\hat s-4m_s^2}}
 \left[
 \frac{18\lambda_s^2u^2(\hat s-4M^2)}{(\hat s-m_s^2)^2}
 -\frac{24\lambda\lambda_suM}{\hat s-m_s^2}
 -\frac{\lambda^2\left(2\hat sM^2+16M^4-16M^2m_s^2+3m_s^4\right)}{\hat sM^2-4M^2m_s^2+m_s^4}
 \right.\nonumber\\
 &\left.
 +\lambda\left\{\frac{\lambda\left(\hat s^2+16\hat sM^2-32M^4+6m_s^4-4\hat sm_s^2-16M^2m_s^2\right)}{\left(\hat s-2m_s^2\right)\sqrt{\hat s-4m_s^2}\sqrt{\hat s-4M^2}}
 -\frac{12\lambda_suM\left(\hat s-8M^2+2m_s^2\right)}{\left(\hat s-m_s^2\right)\sqrt{\hat s-4m_s^2}\sqrt{\hat s-4M^2}}\right\}\right.\nonumber\\
 &\left.
 \times\log\left(\frac{\hat s-2m_s^2+\sqrt{\hat s-4m_s^2}\sqrt{\hat s-4M^2}}{\hat s-2m_s^2-\sqrt{\hat s-4m_s^2}\sqrt{\hat s-4M^2}}\right)
 \right],
\end{align}
}


where  the symmetry factors of $1/2$ for the initial and final states have been included directly in the cross section. To get $4F \sigma^\mathrm{CM}_{ss\to\nu\nu}$, one uses
\begin{equation}
F={1\over 2} \sqrt{\hat s} \sqrt{\hat s-4 m_s^2} \;,
\end{equation}
which holds  for $ss\rightarrow X$ processes. 

We find good numerical agreement with the corresponding CalcHEP \cite{Belyaev:2012qa} result.

    \subsubsection{$s\rightarrow \nu\nu$ }
    
   This process is allowed when $m_s >2M$.  
  The calculation of the decay rate $s\rightarrow \nu\nu$ is straightforward with the result
     \begin{equation}
 \Gamma(s\rightarrow \nu\nu) = {\lambda^2 m_s \over 16 \pi} \left(   1- {4M^2\over m_s^2}\right)^{3/2} \;.
\end{equation}
The consequent reaction rate is  
    \begin{equation}
 \Gamma_{12}(s\rightarrow \nu\nu) =  {  \Gamma(s\rightarrow \nu\nu)  \;m_s^3\over 2 \pi^2} \int_1^\infty dx {\sqrt{x^2-1} \over e^{{m_s \over T}x}  -1} \;.
 \label{Gamma12}
 \end{equation}

 
 We note that, in the reaction $hh \rightarrow s \rightarrow \nu\nu$, the intermediate $s$ can be on--shell at temperatures below a certain threshold. This reaction corresponds to production   and decay
 of real $s$ included in $ \Gamma_{12}(s\rightarrow \nu\nu) $. To avoid double counting \cite{Weldon:1983jn}, we cut out this resonant region in the $\Gamma_{22}$ rate integral, although
 the result is barely affected.

 \subsection{Dark matter abundance: $m_s >2M$}
 \label{sec:abundance}

 In this subsection, we solve the Boltzmann equation for the neutrino number density and find parameter regions consistent with the observed DM abundance.
 Here we assume that $m_s> 2M$ such that the decay mode $s\rightarrow \nu\nu$ is available. 
 Note that the thermal correction to $M$ is suppressed by  $\lambda^2$ 
 and can therefore be neglected.

 \subsubsection{Qualitative behaviour of the Boltzmann equation solution}
 
 Consider freeze--in production of $N$ particles in the reaction $i \rightarrow N$. In the relativistic regime, the reaction rate scales as $T^{l}$, where $l$ depends on the interaction type.
 Using entropy conservation $g_{*s}a^3 T^3 =$ const with $a$ being the scale factor, one can trade the time variable for $T$.
 The resulting Boltzmann equation can be written as
 \begin{equation}
 T {dn \over dT} -3n + c T^{l-2}=0 \;,
 \end{equation}
where 
 \begin{equation}
 c\equiv {N \; \Gamma (i \rightarrow N) \over H T^{l-2}} 
 \end{equation}
 and we have  taken the number of d.o.f. to be  constant in the range of interest.
 Assuming that the initial density is zero at temperature $T_0$, the solution reads
 \begin{equation}
  n(T) = {c \over 5-l}\; T^3 \; \Bigl(T^{l-5}- T_0^{l-5}\Bigr) \;,
   \end{equation}
   while for $l=5$ it is $n(T)=cT^3 \ln {T_0\over T}$.
 For renormalizable interactions,  $l \leq 4$ and the result at late times is insensitive to $T_0$:
 \begin{equation}
  n(T) \simeq {c \over 5-l}\; T^{l-2} \;.
   \end{equation}
On the contrary, non--renormalizable interactions lead to the ``UV freeze--in'', i.e. the density dominated by the early time production at $T_0$,
\begin{equation}
  n(T) \simeq {c \over l-5}\; T^{3} \; T_0^{l-5}\;,
   \end{equation}
while for $l=5$ $n(T)=cT^3 \ln {T_0\over T}$.
 
 In our work, the relevant reactions are of the type $1\rightarrow 2$, $2\rightarrow 2$ and $2\rightarrow 1$. Their temperature scaling will be discussed later.
 
 \subsubsection{Results}
 
 The Boltzmann equation describing evolution of the $\nu$ number density reads
  \begin{eqnarray}
 \dot{n}_\nu +3n_\nu H &=& 2 \hat \Gamma_{12} (s\rightarrow \nu\nu) + 2 \hat \Gamma_{12} (h\rightarrow \nu\nu) \nonumber\\
 &+&
 2\hat \Gamma_{22} (ss \rightarrow \nu\nu) + 2\hat \Gamma_{22} (hh \rightarrow \nu\nu)  
 + 2\hat \Gamma_{22} (hs \rightarrow \nu\nu)\; .
\end{eqnarray}
 Here 
 \begin{eqnarray}
 && \hat\Gamma_{12} (s\rightarrow \nu\nu) =  \theta(T-T_c^u)  \; \Gamma_{12} (s\rightarrow \nu\nu)\Bigl\vert_{u=0} 
 + \; \theta(T_c^u-T) \; \Gamma_{12} (s\rightarrow \nu\nu) \;, \\
 && \hat\Gamma_{12} (h\rightarrow \nu\nu) =   \theta(T_c^u-T)   \;  \theta(T_c^v-T)    \; \Gamma_{12} (h\rightarrow \nu\nu) \;, \\
 && \hat\Gamma_{22} (ss\rightarrow \nu\nu) =  \theta(T-T_c^u)  \; \Gamma_{22} (ss\rightarrow \nu\nu)\Bigl\vert_{u=0} 
 + \; \theta(T_c^u-T) \; \Gamma_{22} (ss\rightarrow \nu\nu) \;, \\
 && \hat \Gamma_{22} (hh \rightarrow \nu\nu) = \theta(T_c^u-T) \; ( 4 - 3 \theta(T_c^v-T) \;\Gamma_{22} (hh \rightarrow \nu\nu) \;,\\
 && \hat \Gamma_{22} (hs \rightarrow \nu\nu) = \theta(T_c^v-T)  \;\Gamma_{22} (hs \rightarrow \nu\nu) \;.
 \end{eqnarray}
 The theta--functions make sure that the processes involving scalar trilinear vertices are switched off above the critical temperatures. Further, they take care of the 
 different number of Higgs d.o.f. before and after electroweak phase transition. The rates $\Gamma_{22}$ and $\Gamma_{12}$ are calculated according to
 (\ref{22rate}) and (\ref{Gamma12}) using the results of the previous subsections with non--zero $v$ and $u$. 
 The Higgs decay rate $\Gamma_{12} (h\rightarrow \nu\nu)$   is given by $\sin^2 \theta \; \Gamma_{12} (s\rightarrow \nu\nu)$.
 
 The Boltzmann equation in the relativistic regime has a simple solution. We find  that the most important contribution comes from  $\Gamma_{12}$. Since $\lambda_{hs} $
 and $\lambda_s$ are small, we may neglect the $s$--thermal mass at late times, in which case Eq.\;\ref{Gamma12} yields
 \begin{equation}
 \Gamma_{12} \propto m_s^2 T^2 \;,
 \end{equation} 
 while at very high temperatures it scales as   $ m_s^4(T) \propto T^4$. In this regime,
 \begin{equation}
 n_\nu(T) \simeq {\rm const} \;,
 \label{analytic-sol}
 \end{equation}
   where the constant is proportional to $m_s^2$.
 The dark matter yield is conveniently expressed in terms of $Y$,
  \begin{equation}
Y= {n_\nu\over s_{\rm SM}} ~~,~~ s_{\rm SM}= {2\pi^2 g_{*s}\over 45} \; T^3 \;,
\end{equation}
 where $g_{*s}$ is the number of d.o.f. contributing to the entropy. 
 It is proportional to the total number of the DM quanta.
 The observed DM density requires
  \begin{equation}
Y_\infty=  4.4\times 10^{-10} \; {\left( {\rm GeV}\over M \right)} \;.
\end{equation}
The solution (\ref{analytic-sol}) is valid in the relativistic regime, that is, down to temperatures of order $m_s$.
Thus, the resulting  $Y \propto 1/m_s$.
 
 \begin{figure}[h!]
\begin{center}
\includegraphics[scale=0.65]{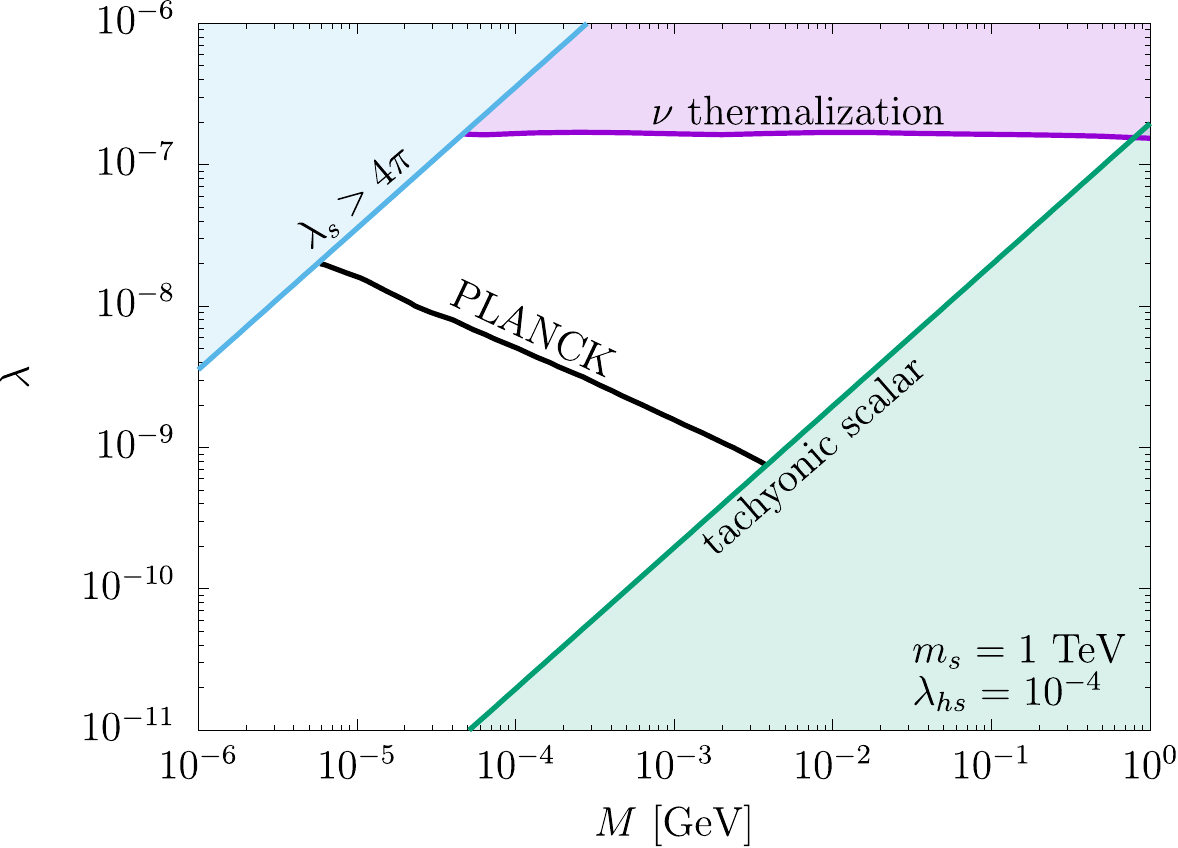}
\includegraphics[scale=0.65]{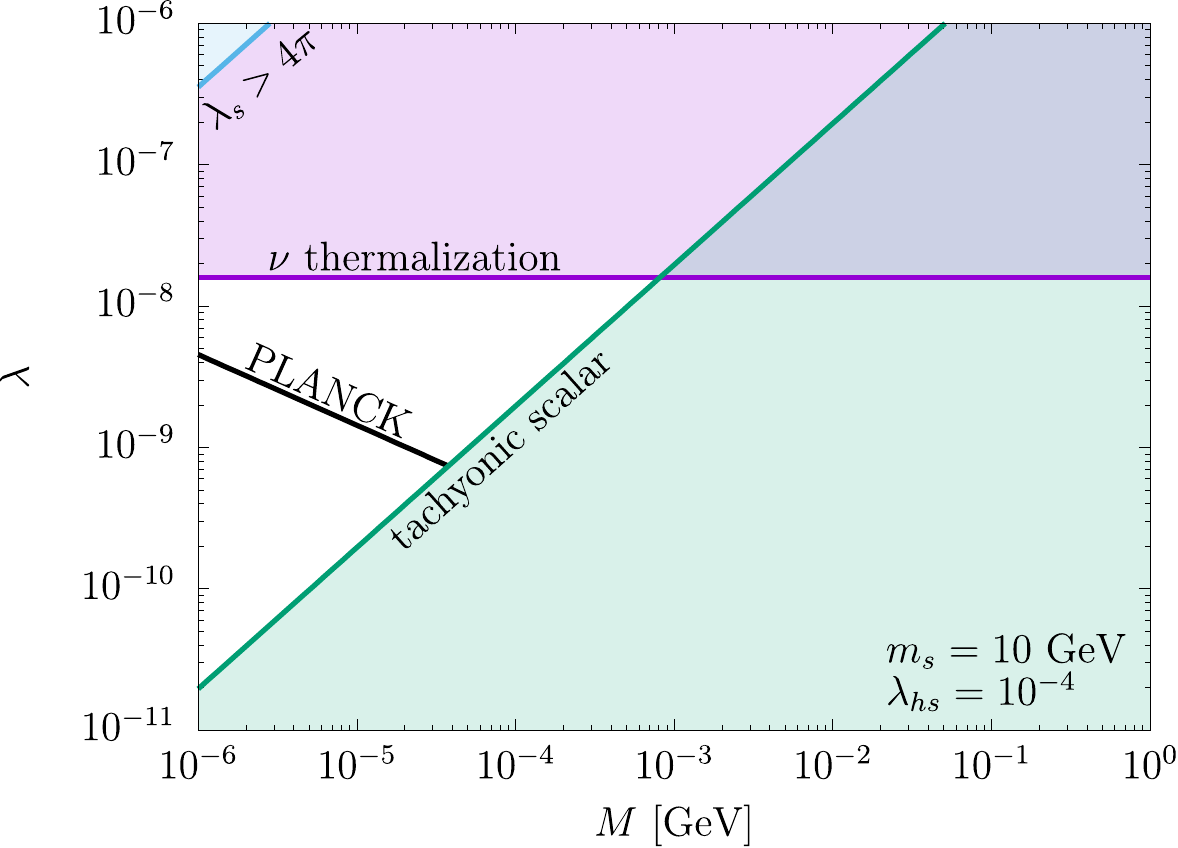}
 \includegraphics[scale=0.65]{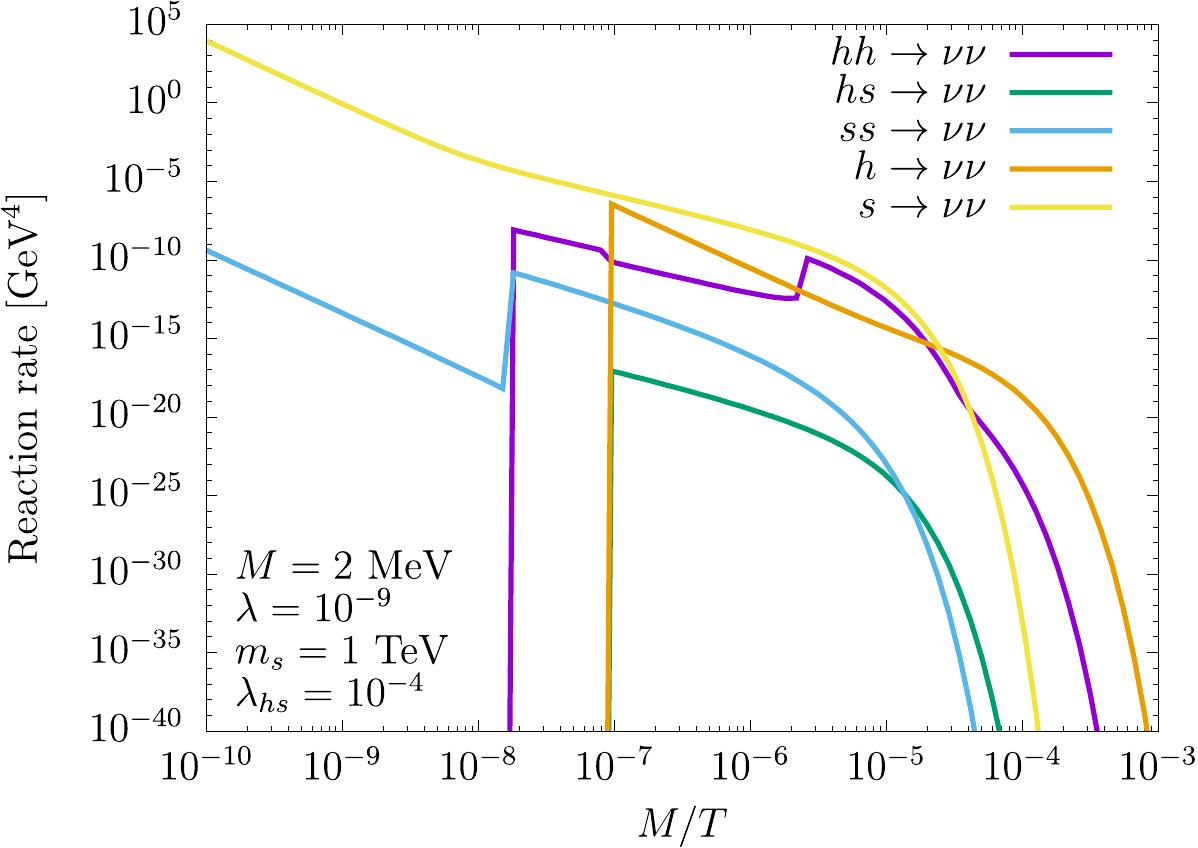}
 \includegraphics[scale=0.65]{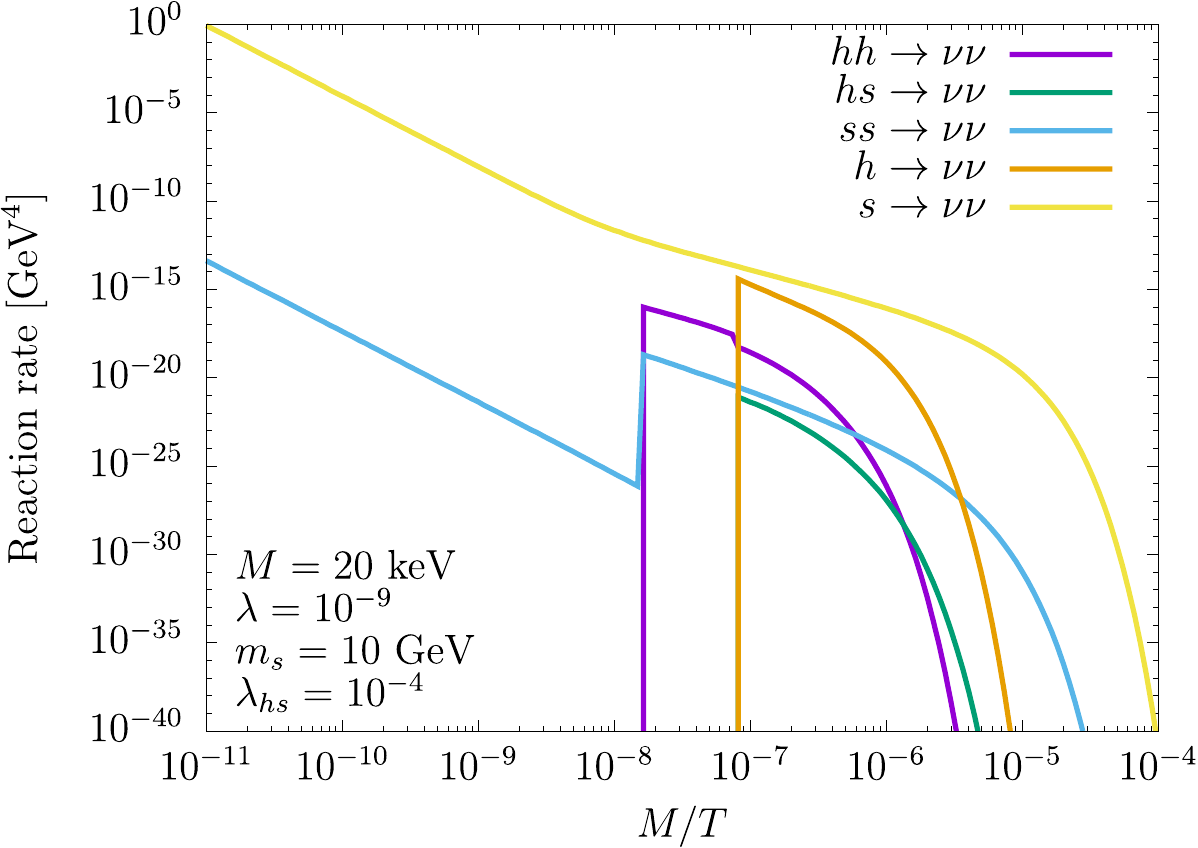}
 \end{center}
\caption{   {\it Upper panels:} $\lambda$ vs $M$ producing the correct DM relic density (``PLANCK''). The dominant DM production mode is $s \rightarrow \nu\nu$.
{\it Lower panels:} Reaction rates. The kinks appear due to phase transitions which bring in new modes as well as due to $T$--dependent   propagators.
}
\label{planck}
\end{figure}

Our numerical results for the total DM relic abundance and the full reaction rates are shown in Fig.\;\ref{planck}. 
We find that the DM yield is dominated by the decay $s\rightarrow \nu\nu$ at temperatures $T \sim m_s$
and the required coupling is
\begin{equation}
\lambda \simeq 1.7 \times 10^{-12}\; \sqrt{ m_s \over M} \;.
\label{lambda-ms-M}
\end{equation}
This applies to the regime $m_s \gg M$. In this case, the DM yield $Y$ due to the  $s\rightarrow \nu\nu$ decay is independent of $M$ and 
proportional to $\lambda^2$. Thus, in order to get the right relic abundance, the relation $\lambda \propto 1/\sqrt{M}$ is enforced (while smaller $M$ for the same 
$\lambda$ lead to under-abundance).
We find that these conclusions apply quite generally, beyond the parameter choices of Fig.\;\ref{planck}.  

Given the correct relic abundance, small and large values of $M$ are excluded by perturbativity and the Higgs mixing or the presence of a tachyonic scalar.
Indeed, since $\lambda u =M$ and $m_s^2 = 2 \lambda_s u^2$,
\begin{equation}
\lambda_s = {\lambda^2 m_s^2 \over 2 M^2} \;.
\end{equation}
For a fixed relic density and other parameters, $\lambda_s \propto 1/M^3$ so that at low $M$ it blows up while for large $M$ it violates 
$4 \lambda_h \lambda_s > \lambda_{hs}^2$.  

Since our focus is on {\it freeze--in} production of neutrino DM, we exclude significant values of $\lambda$. These lead to efficient $\nu$ production such that $n_{\nu}$ is close to its equilibrium value. In this  case, the reverse process $\nu\nu \rightarrow s$ becomes important and the system tends to equilibrate. Although such a possibility is not excluded by observations, it does not correspond to freeze--in neutrino production.

The approximation  $\theta \ll 1$ applies in all of the allowed parameter space: $\theta$ ranges from $10^{-1}$ in the lower right corner to $10^{-5}$ in the upper left corner of the plots.
Close to the tachyonic region however, $\theta \sim m_h/m_s$ or $m_s/m_h$ such that the relations (\ref{appr}) receive non--negligible corrections. Therefore the tachyonic region border is only approximate.

The stability condition $4\lambda_s \lambda_h > \lambda_{hs}^2$ combined with the right DM yield $Y$ impose a lower bound on $m_s$,
\begin{equation}
m_s >   10^8 \;  \lambda_{hs}^{2/3} M  > 1 ~ {\rm MeV} \;.
\end{equation}
To get the limit of 1 MeV, we have used  $\lambda_{hs}> 4 \times 10^{-8}$ required for thermalization  and  the warm DM bound $M> 1$ keV   (taking the number of SM degrees of freedom at $m_s$ to be 10).\footnote{The exact lower bound on warm dark matter mass  from free streaming  \cite{Irsic:2017ixq} depends on its momentum distribution. See, e.g. 
 \cite{Kamada:2019kpe,Huo:2019bjf}  for recent analyses. }

  \begin{figure}[h!]
\begin{center}
\includegraphics[scale=0.65]{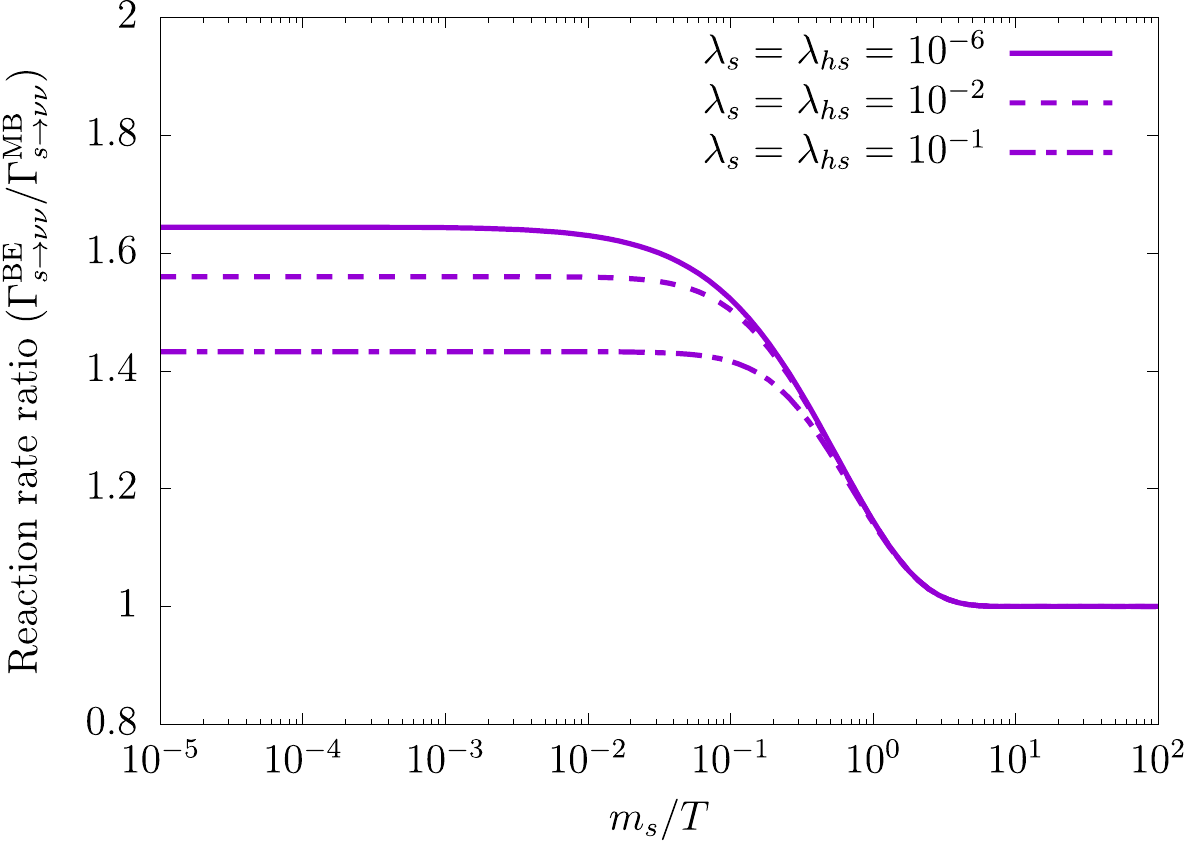}
 \end{center}
\caption{     Bose--Einstein vs Maxwell--Boltzmann $s \rightarrow \nu \nu$ reaction rates.}
\label{ratio}
\end{figure}

The main DM production channel is $s\rightarrow \nu\nu$. We find that the relativistic effects in this reaction are tangible. Fig.\;\ref{ratio}
shows that replacing the Bose--Einstein distribution with the Maxwell--Boltzmann one can lead to up to a 65\% error in the reaction rate.
The Bose--Einstein enhancement is sensitive to the thermal mass: for lower couplings the effect is more pronounced. This is natural since 
the distribution peaks at low energies while the thermal mass provides a lower bound on how low the energy can be.

\subsection{Light $s$: $m_s<2M$ }

In this case, the main production channel $s\rightarrow \nu\nu$ becomes less significant. The process is kinematically allowed at very high temperatures, when $u=0$ and the Majorana neutrino mass vanishes. It stops after the phase transition to $u\not= 0$.
 The produced number density is    diluted by the subsequent Universe expansion.  As a result, the processes like $ss\rightarrow \nu\nu$ and $h\rightarrow \nu\nu$
become equally important or even 
take over the leading role.

 \begin{figure}[h!]
\begin{center}
\includegraphics[scale=0.65]{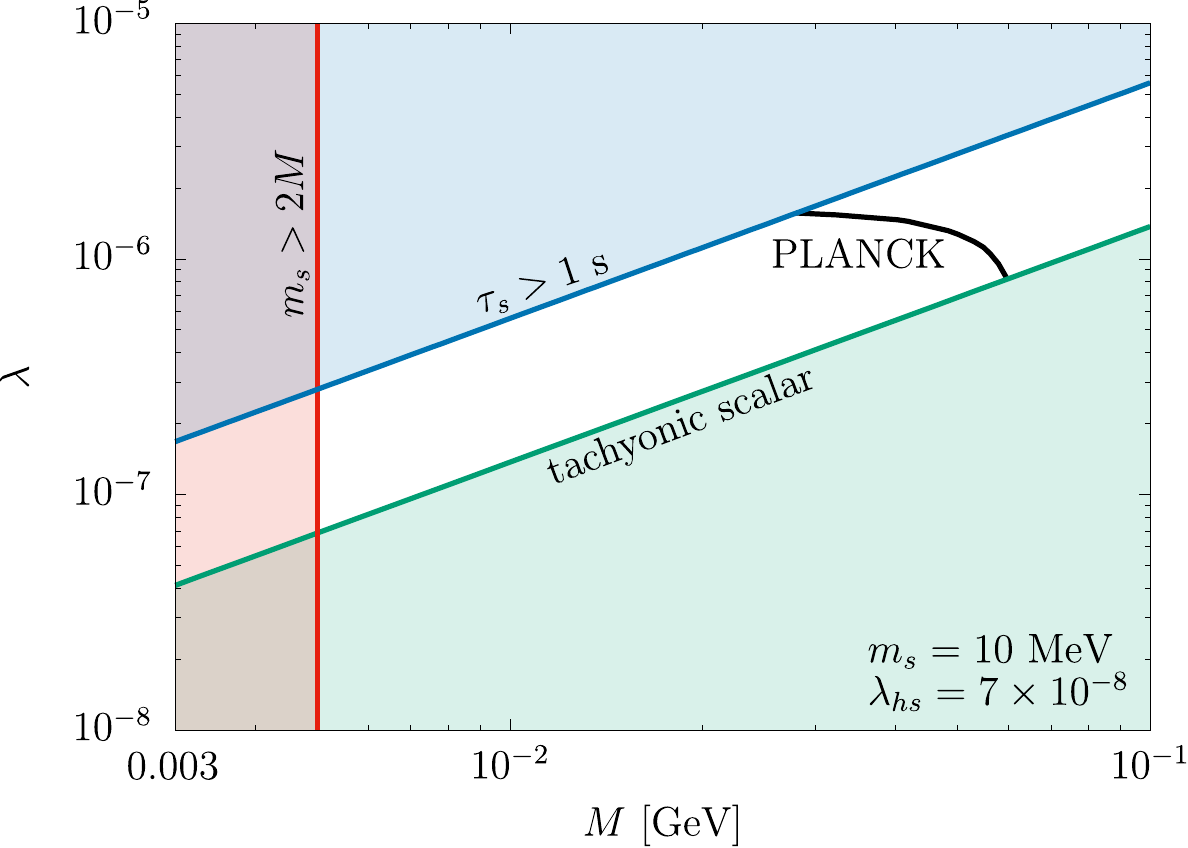}
\includegraphics[scale=0.65]{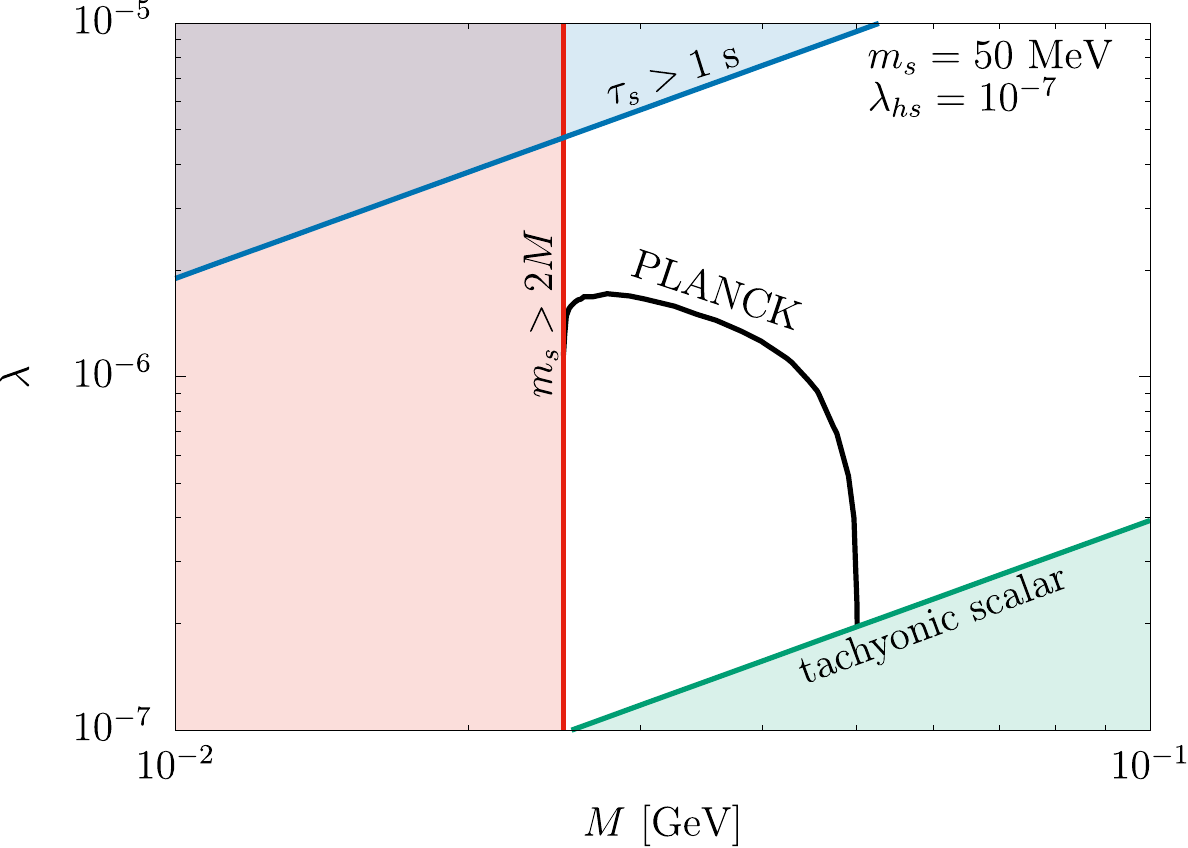}
 \includegraphics[scale=0.65]{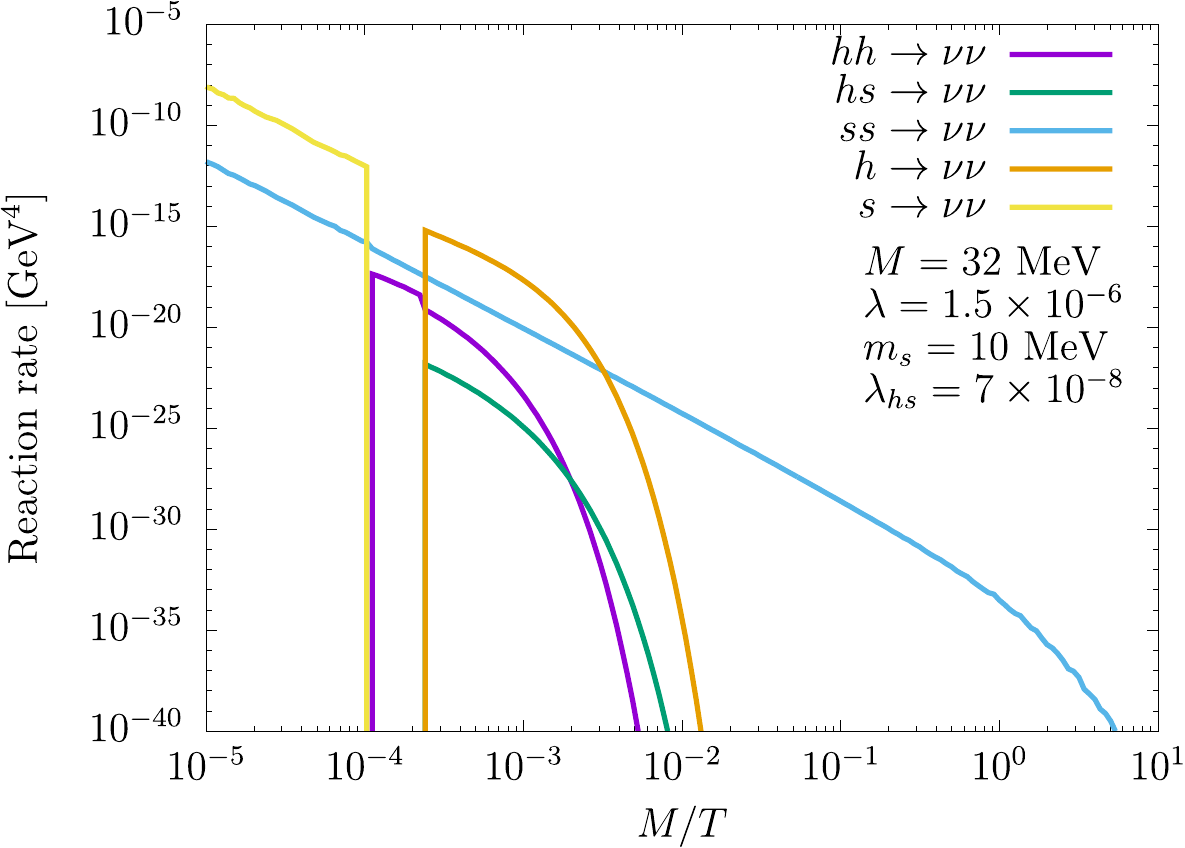}
 \includegraphics[scale=0.65]{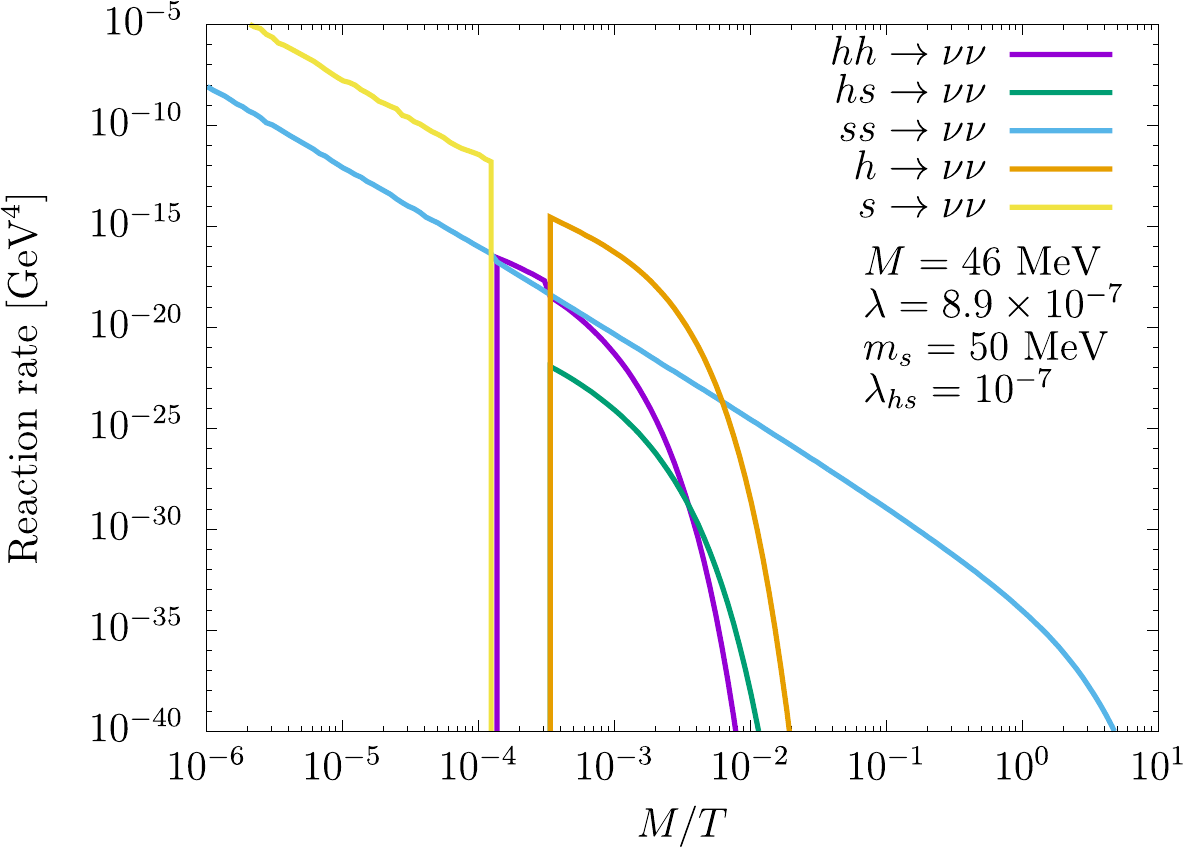}
 \includegraphics[scale=0.65]{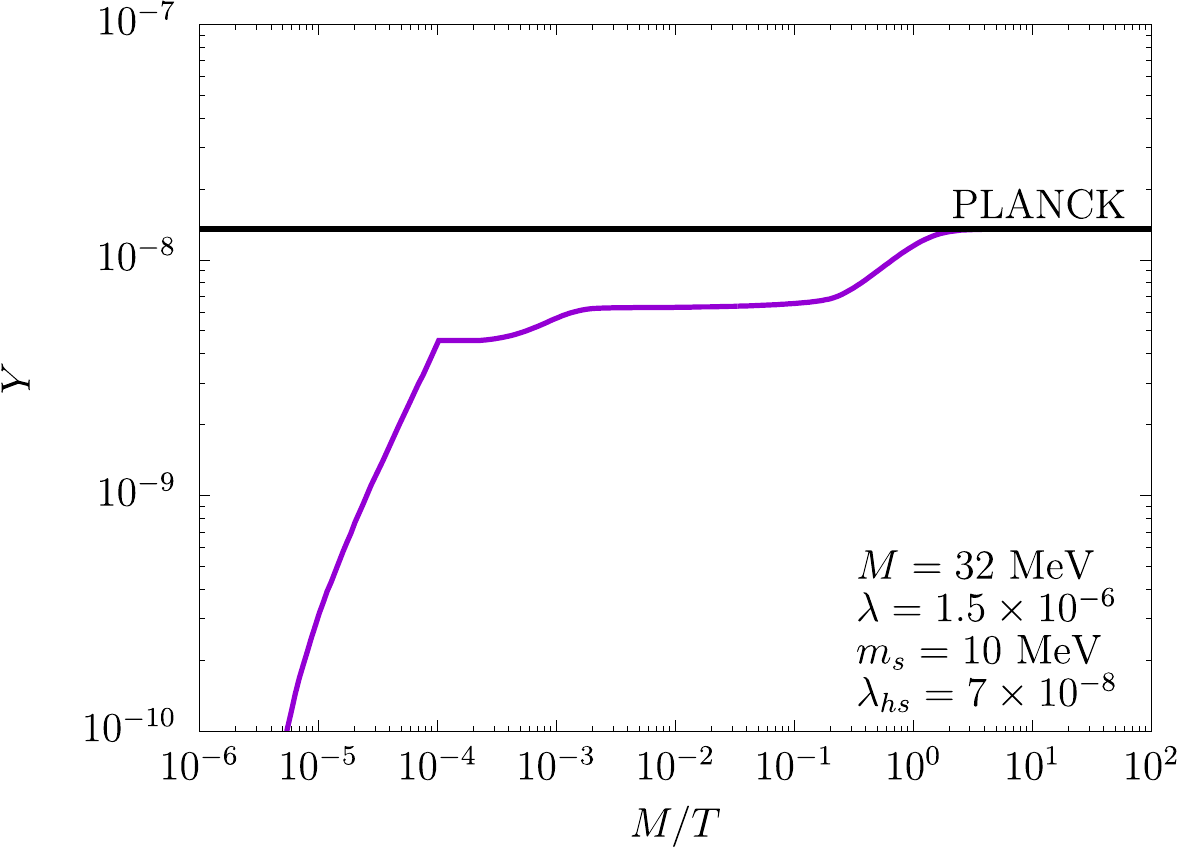}
 \includegraphics[scale=0.65]{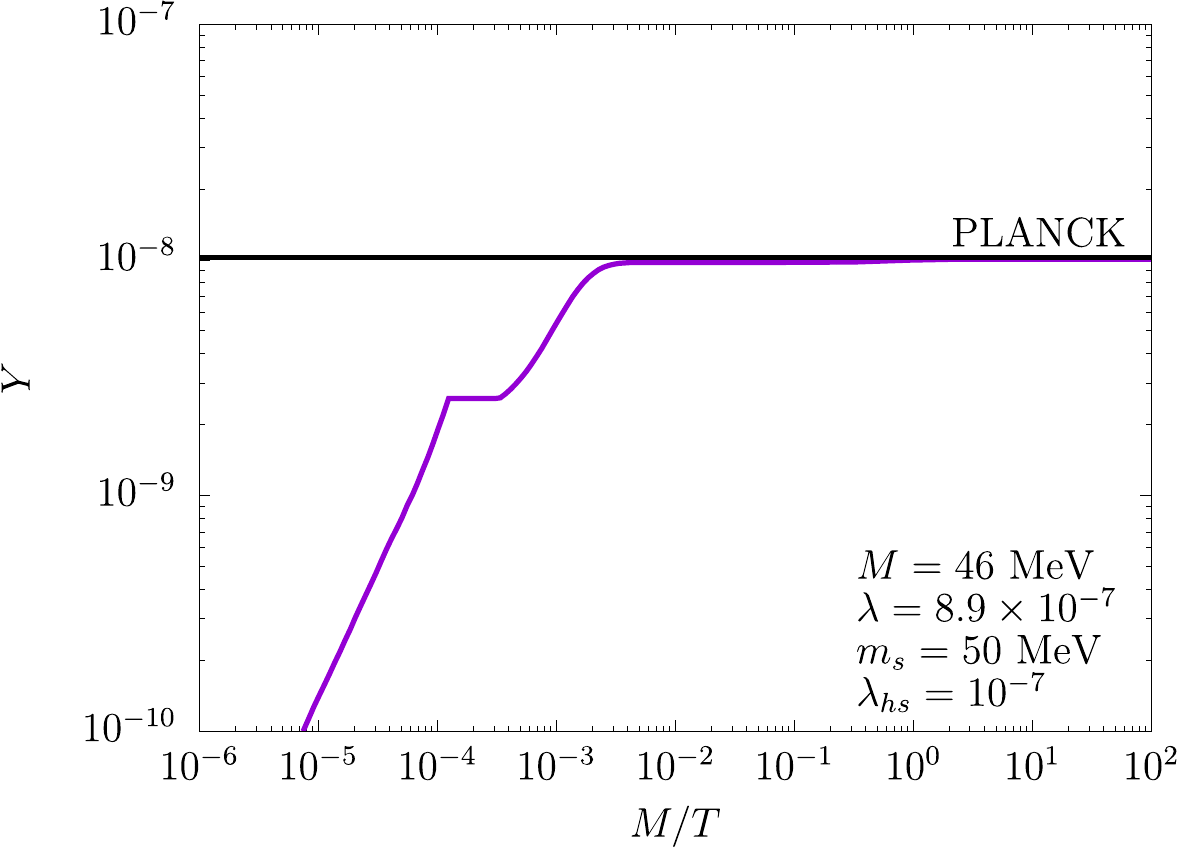}
 \end{center}
\caption{   {\it Upper row:} $\lambda$ vs $M$ producing the correct DM relic density (``PLANCK'') for $m_s < 2M$. 
{\it Middle row:} Reaction rates for  representative parameter sets. 
{\it Lower row:} DM yield for the above parameter sets. Left: $ss\rightarrow \nu\nu$  and  $s\rightarrow \nu\nu$ dominate. Right:  $h \rightarrow \nu\nu$ dominates.}
\label{planck-light-s}
\end{figure}

 In case of a very light $s$, there are a number of non--trivial constraints to be observed. In particular, one must make sure that $s$ decays before BBN.  Since $s$ cannot decay into neutrinos, the decay proceeds through the mixing with the Higgs. The decay modes and widths are discussed in Appendix B. We impose the  constraint
  $\tau < 1$sec, which ensures that $s$ does not contribute to the relativistic degrees of freedom at BBN and does not destroy light nuclei. Furthermore,
  for $m_s <2 m_\mu$, there is a strong constraint on the mixing angle with the Higgs.  
Rare Kaon decays require $\theta \lesssim 10^{-4}$  \cite{Andreas:2010ms}. For heavier $s$, the bound relaxes to  $10^{-3}$ or so  \cite{Schmidt-Hoberg:2013hba}.
Finally, since we are assuming a thermal abundance for $s$, the Higgs portal coupling must be large enough
to ensure thermalization via $h \leftrightarrow ss$,
$\lambda_{hs} > 4\times 10^{-8}$.

Although the available parameter space is quite limited, we find that it is still possible to obtain  the right DM relic density. Two examples are 
 shown in Fig.\;\ref{planck-light-s}.
In this case, the strongest constraints are imposed by $\tau_s <1$ sec and the absence of tachyons, $4\lambda_h \lambda_s > \lambda_{hs}^2$. The latter is 
significant since a light $s$ requires a small $\lambda_s$. In the allowed parameter space, the bound $\theta \lesssim 10^{-4}$ is then satisfied.

As seen in the plots, different reactions dominate at different times. At high temperatures, $s\rightarrow \nu\nu$ dominates but the resulting DM density gets diluted. At later times,
$h \rightarrow \nu\nu$ and $ss\rightarrow \nu\nu$ become important. The  plateau regions producing the correct DM relic density (Fig.\;\ref{planck-light-s}, upper row) are 
associated with $ss\rightarrow \nu\nu$  as the leading (or next-to-leading) production mode. The corresponding rate scales as $T^4$ down to temperatures of order $M$. Thus, the resulting
yield  satisfies
\begin{equation}
Y \propto {1\over M} \;.
\end{equation}
Since the required $Y_\infty$ also scales as $1/M$, the PLANCK line corresponds to a plateau in the $(M,\lambda)$ plane. The DM yield associated with the different reactions is shown in the lower row of
Fig.\;\ref{planck-light-s}. The left panel confirms that more than 50\% of the yield in  the plateau region is indeed provided by $ss\rightarrow \nu\nu$.   
We also observe that $s\rightarrow \nu\nu$ makes a significant  contribution and tilts the PLANCK line in analogy with Fig.\,\ref{planck}.

At somewhat larger masses, the Higgs decay $h \rightarrow \nu\nu$ becomes more important. The amplitude for this process is proportional to $\lambda \, \theta$ which is approximately constant for a fixed $M$:
\begin{equation}
\lambda \, \theta \simeq {\lambda_{hs} v M \over m_h^2} \;.
\end{equation}
Thus, the resulting PLANCK region is almost vertical in the $(M,\lambda)$ plane. The lower right panel of Fig.\;\ref{planck-light-s} shows that the dominant DM yield is 
produced at electroweak temperatures via  $h \rightarrow \nu\nu$. To the left of the PLANCK curve, our DM is under-abundant.

The neutrino thermalization constraint of Fig.\;\ref{nu-therm} is not directly applicable here since the channel $s \rightarrow \nu\nu$ is not available. 
We find that, in the allowed parameter region, $n_\nu$ is below its equilibrium value so the neutrinos can be treated as non--thermal.

\section{Sterile neutrino production II: non--thermal $s$}
\label{secII}

It is possible that $s$ never reaches thermal equilibrium either due to its large mass or due to its small couplings.
In general, there is a variety of non--thermal $s$--production mechanisms in the Early Universe. Its direct coupling to an inflaton would lead to perturbative and/or non--perturbative production, e.g.  via parametric resonance \cite{Kofman:1994rk}. Furthermore, light scalar field fluctuations during inflation generate an $s$--condensate which then decays into $s$--quanta. However, these mechanisms are sensitive to further details of the complete UV model, for instance, to the Hubble rate during inflation  \cite{Kainulainen:2016vzv}. In particular, for small Hubble rates
such contributions are suppressed. 
In what follows, we focus on $s$--production from a Standard Model thermal bath and assume that the other sources are subdominant.

\subsection{Heavy $s$}

If $s$ is very heavy while the temperature is not high enough, the singlet does not thermalize and can be integrated out. DM production proceeds 
through Higgs annihilation $hh \rightarrow \nu\nu$ and decay $h \rightarrow \nu\nu$ due to the Higgs--singlet mixing. 
We find that the decay  mode dominates for the parameter values of interest. 

It is instructive to consider the channel $hh \rightarrow \nu\nu$ separately. When this mode dominates, one recovers the so--called ``UV freeze--in'' scenario.
 In this  case, the DM abundance is sensitive to the maximal temperature $T_0 < T_c^u$.
  The Boltzmann equation at high $T$ reads
 \begin{equation}
 T {dn \over dT} -3n + {8 \Gamma_{22}(hh\rightarrow \nu\nu) \over H}=0 \;,
\end{equation}
where the factor of 8 takes into account 4 Higgs d.o.f. above the EW transition scale.
Since $\Gamma_{22} \propto T^6$ at high $T$, 
 \begin{equation}
 n(T) \propto T_0T^3\;.
\end{equation}
 As a result, the DM yield $Y = n/s_{\rm SM} \propto  T_0$ is determined by the UV end of the evolution. This is unlike the usual freeze--in scenario where 
  the  IR behaviour is more important.
 
\begin{figure}[h!]
\begin{center}
\includegraphics[scale=0.65]{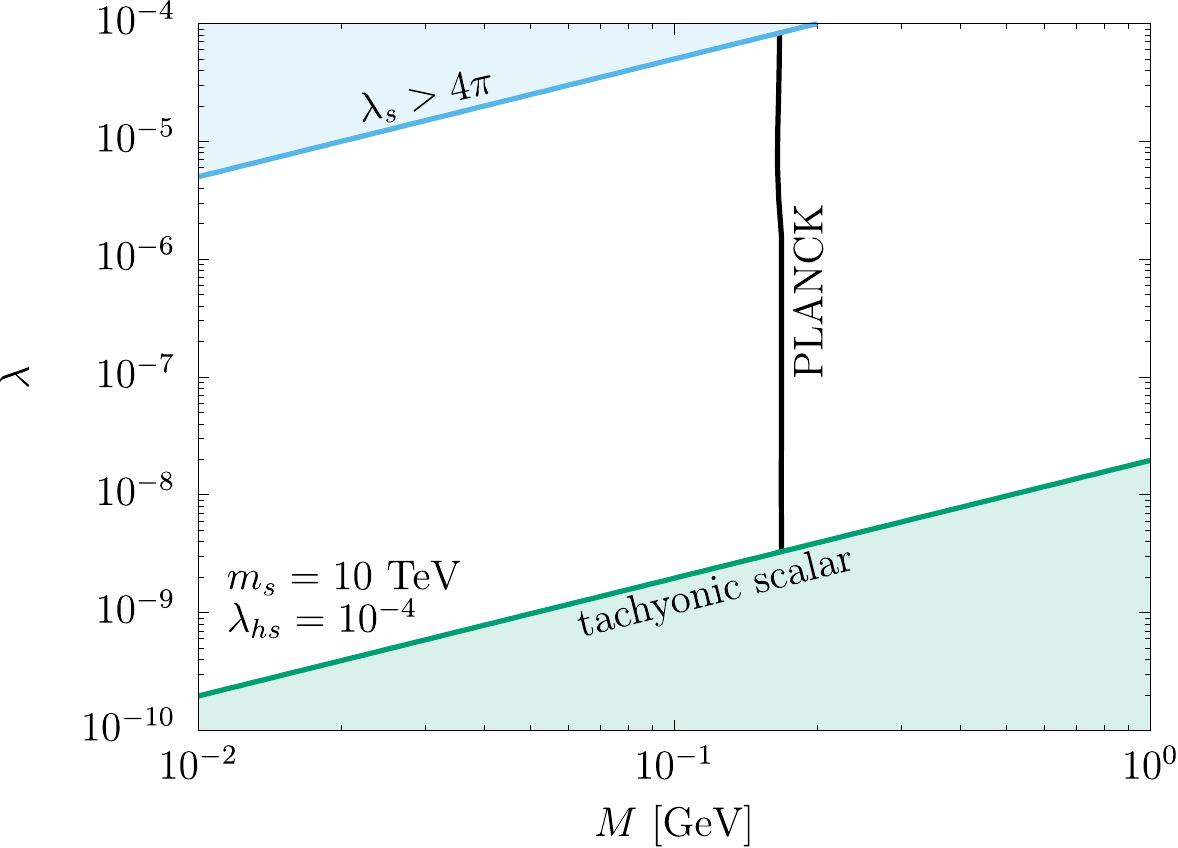}
\includegraphics[scale=0.65]{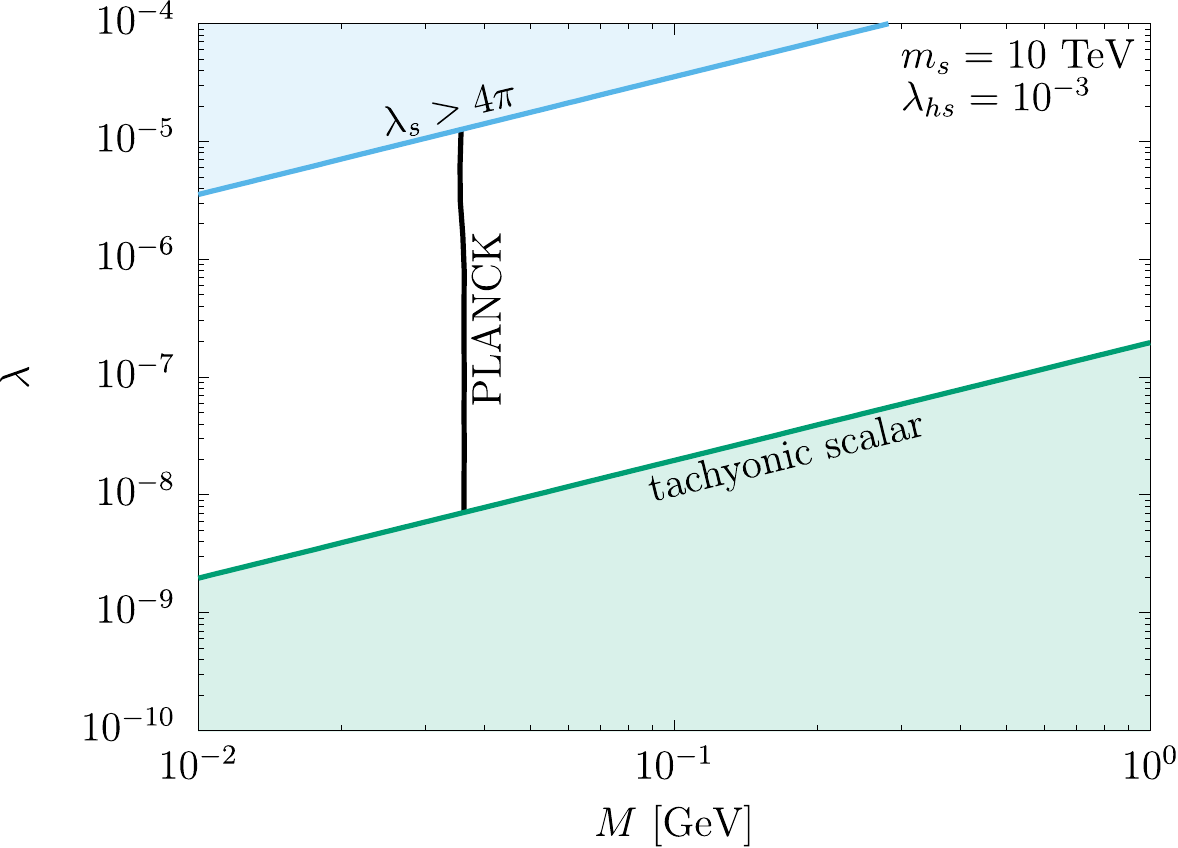}
 \end{center}
\caption{      $\lambda$ vs $M$ producing the correct relic DM relic density (``PLANCK'') for a heavy $s$. The dominant DM production mode is $h \rightarrow \nu\nu$.
The maximal temperature is chosen to be $T_0=1$ TeV. }
\label{T0}
\end{figure}

 Although the decay channel $h \rightarrow \nu\nu$ opens up only below the EW breaking scale, numerically it turns out to be more important and the sensitivity of the DM abundance to  $T_0$ is  weak. 
 Our numerical results are presented in Fig.\;\ref{T0} which shows the regions with the right  relic abundance. The DM production amplitude  is proportional to the combination $\lambda \theta$  which  
 is fixed for a fixed $M$. This makes the production rate independent of $\lambda $ and the PLANCK region vertical in the $(M, \lambda)$ plane.
 As before, our DM is under-abundant to the left of the PLANCK line.

The Higgs portal coupling required for the correct DM relic abundance can be approximated by (ignoring the phase transition complications):
\begin{equation}
\lambda_{hs} \simeq   \; {m_s^2 \over  M^{3/2}} ~ {4 \times 10^{-14} \over \sqrt{\rm GeV}}
\end{equation}
for $g_* \simeq 107$. We have verified that the neutrino thermalization constraint is insignificant and $n_\nu$ is below its equilibrium value. 

In the allowed parameter space, the mixing angle ranges from $10^{-2}$ to $10^{-6}$. As before, the $\theta^2$ corrections become significant close to the tachyonic region border.

\subsection{Small couplings: freeze--in production of $s$}
\label{freeze-in-s}

Here we consider the possibility that the $\lambda_{hs} $ and $\lambda_s$ couplings are so small that $s$ never reaches thermal equilibrium
(see, e.g. \cite{Heikinheimo:2016yds} for early work). Assuming that the initial abundance of $s$ is zero or negligibly small, the $s$ quanta are produced by the Higgs thermal bath via the usual freeze--in mechanism. Subsequently, they decay into sterile neutrinos 
leading to the required DM abundance. Due to the $s-h$ mixing, $s$ decays also produce SM particles, yet this gives only a small correction to the entropy since the 
density of $s$ is far below its equilibrium value.  

There are a few $s$--production channels: $hh \rightarrow ss$, $h \rightarrow ss$ and $hh\rightarrow s$, where the last two reactions are possible only 
below the corresponding critical temperatures. 
 $hh \rightarrow s$ is a new reaction type, not considered before. Hence, it is instructive to consider it in more detail.

\subsubsection{$h h \rightarrow s$ rate}

The general expression for the reaction rate reads
\begin{equation}
\Gamma_{2 1} = \int \left( \prod_{i\in a} {d^3 {\bf p}_i \over (2 \pi)^3 2E_{i}} f(p_i)\right)~
  {d^3 {\bf p}_f \over (2 \pi)^3 2E_{f}} 
\vert {\cal M}_{2\rightarrow 1} \vert^2 ~ (2\pi)^4 \delta^4(p_1+p_2-p_f) . 
\label{Gamma}
\end{equation}
Here $ \vert {\cal M}_{2\rightarrow 1} \vert^2$ includes $1/2$ from the phase space symmetry of the initial state.

Performing the angular integrals as before  and using 
  \begin{equation}
\int  {d^3 {\bf p_f}\over (2\pi)^3 2E_f} \; (2\pi)^4 \delta(p_1+p_2-p_f) = {\pi \over 2 m_s} \; \delta (E-m_s/2) \;
\end{equation}
 as well as  
  $\vert {\cal M}_{2\rightarrow 1} \vert^2 = 1/2 \times \lambda_{hs}^2 u^2$, we find
  \begin{equation}
\Gamma_{2 1} = {\lambda_{hs}^2 u^2 m_s T \over 32 \pi^3  } \theta(m_s- 2m_h) \int_0^\infty  d\eta {\sinh\eta \over e^{m_s \cosh \eta \over T}-1} \;
\ln {   \sinh { m_s \cosh \eta +  \sqrt{m_s^2-4m_h^2}  \sinh\eta  \over 4T}  \over 
 \sinh { m_s \cosh \eta -  \sqrt{m_s^2-4m_h^2}  \sinh\eta   \over 4T}}  \;.
\end{equation}
  This expression is valid for a single Higgs d.o.f.

  \subsubsection{$ h \rightarrow ss$ and $hh \rightarrow ss$ rates}
  
  These reaction rates have been computed in \cite{Lebedev:2019ton}.  
  For a single Higgs d.o.f., the results read
 \begin{eqnarray}
  \Gamma_{1 2} &=&  { \lambda_{hs}^2 v^2 m_h^2 \over 64 \pi^3   } \sqrt{1-{4m_s^2 \over m_h^2}} 
 \; \int_1^\infty dx \; { \sqrt{x^2-1} \over e^{{m_h\over T} x} -1}  \;, \nonumber \\
  \Gamma_{2 2} &=&  {1\over 2! 2!} \; { \lambda_{hs}^2 T \over 16 \pi^5} \\
&  \times & \int_{m_h}^\infty dE ~E \sqrt{E^2-m_s^2} \int_0^\infty d\eta {    \sinh \eta \over e^{{2E\over T} \cosh\eta }-1}~
\ln {  \sinh    {E\cosh\eta + \sqrt{E^2 -m_h^2} \sinh\eta \over 2T}    \over
\sinh   {E\cosh\eta - \sqrt{E^2 -m_h^2} \sinh\eta \over 2T}  } \;, \nonumber
\end{eqnarray}
where $E $ is   half the CM energy and we have factored out the symmetry factor $1/2!2!$ stemming from 2 identical particles in the initial and final states.

  \subsubsection{Results}
  
  The number density of the $s$--quanta is calculated according to 
  \begin{equation}
 \dot{n}_s +3n_sH = 2 \hat \Gamma_{12} (h\rightarrow ss) + 2\hat \Gamma_{22} (hh \rightarrow ss) + \hat \Gamma_{21} (hh \rightarrow s)  
  \;,
\end{equation}
where
 \begin{eqnarray}
 && \hat\Gamma_{12} (h\rightarrow ss) =    \theta(T_c^v-T) \; \Gamma_{12} (h\rightarrow ss) \;, \\
 && \hat\Gamma_{22} (hh\rightarrow ss) =    ( 4 - 3 \theta(T_c^v-T)) \; \Gamma_{22} (hh \rightarrow ss)   \;, \\
 && \hat \Gamma_{21} (hh \rightarrow s) =   ( 4 - 3 \theta(T_c^v-T)) \;\Gamma_{21} (hh \rightarrow s) \;.
 \end{eqnarray}
 Here the $\theta$--functions account for the EW phase transition and the change in the number of the Higgs d.o.f. 
 We neglect the dependence on $T_c^u$ since $s$ is not thermalized and $\lambda_{hs}$ is very small.
  
  Since there is no significant back reaction of the produced $s$ quanta on the thermal bath nor substantial entropy production via $s$--decay, the total DM yield can then be computed as the $s$--yield times the branching ration for the $s$ decay into dark matter,
  \begin{equation}
 Y_\nu = 2 \;Y_s \; {\rm BR}(s\rightarrow \nu\nu)\;.
\end{equation}
  The $s$ decay width into the SM particles is given in Appendix B.

    \begin{figure}[h!]
\begin{center}
\includegraphics[scale=0.345]{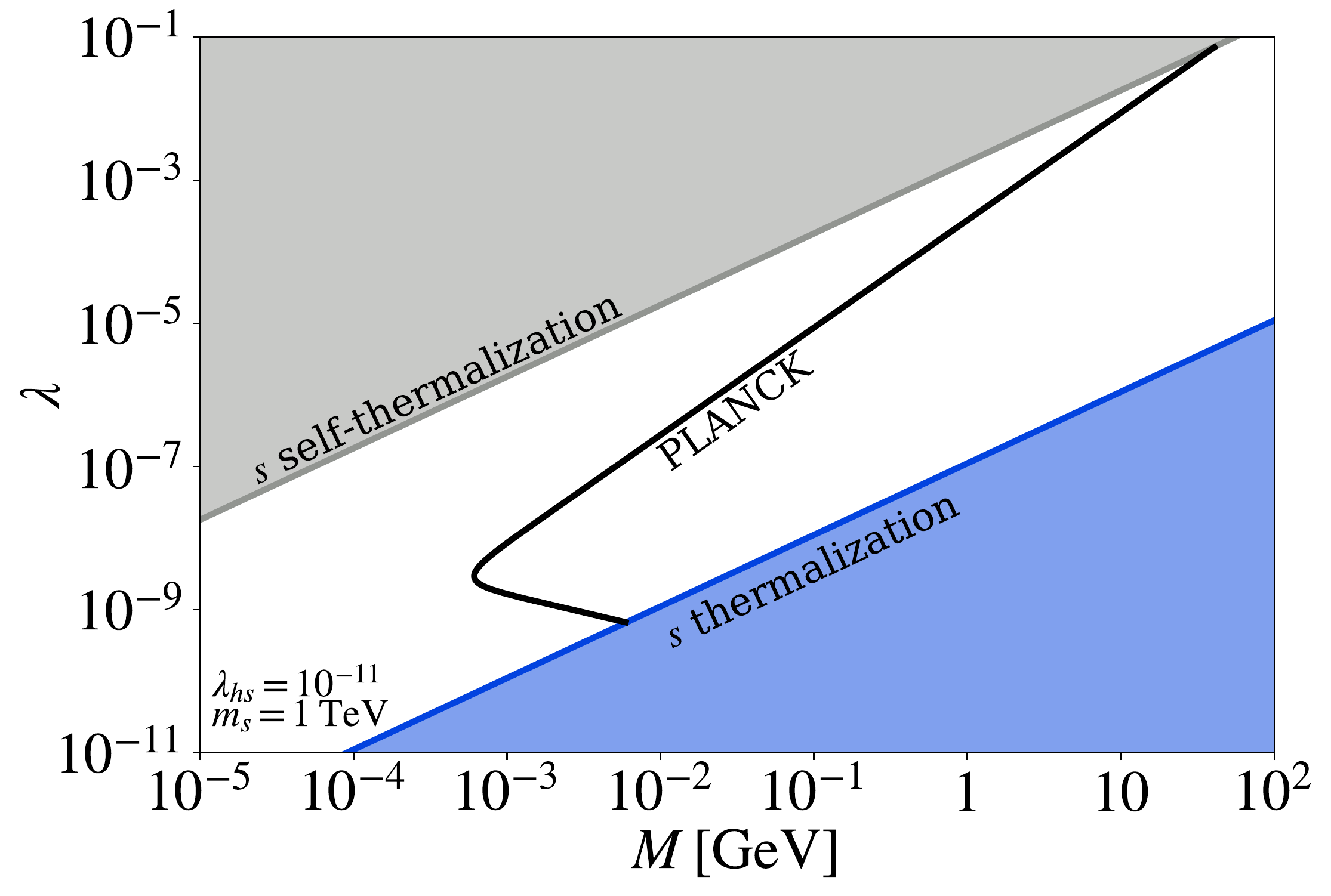}
\includegraphics[scale=0.345]{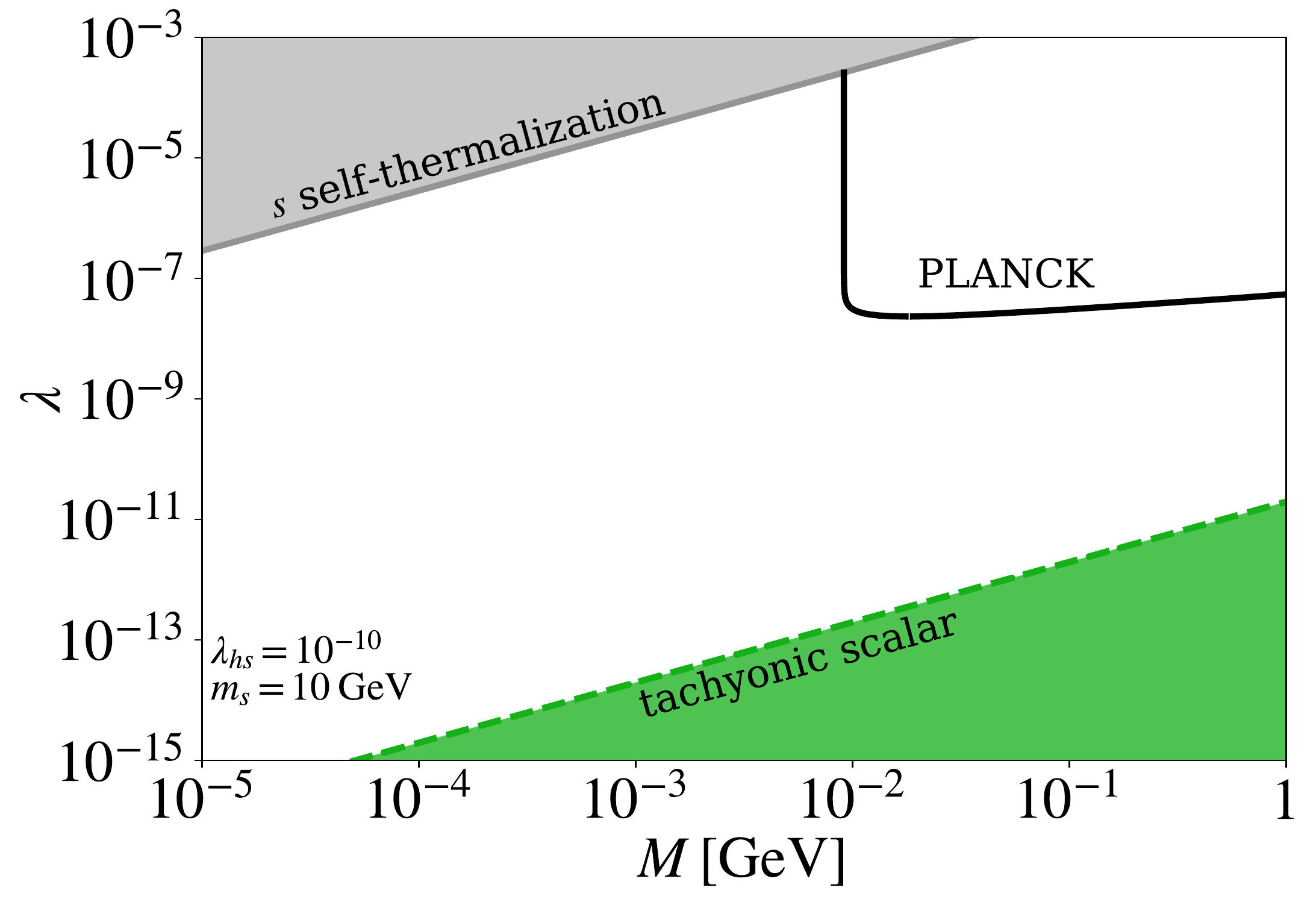}
 \end{center}
\caption{     $\lambda$ vs $M$ generating the correct relic DM relic density (``PLANCK'') for a non--thermal $s$. Here $s$ is produced by the freeze--in mechanism via the Higgs thermal bath. In the excluded regions,
  the freeze--in calculations become unreliable   due to efficient $hh \leftrightarrow s$ or $ss \rightarrow ssss$ processes leading to thermalization.}
\label{planck-non-thermal-s}
\end{figure}

  We compute the number density of $s$ via freeze--in calculations. Thus, it is important to observe the non--thermalization constraints. For a given $\lambda$, an increase in $M$
  implies an increase in $u$, which makes $s$ production  via $hh \rightarrow s$ more efficient and can lead to thermalization. A competitive  constraint, which becomes stronger for light $s$,
  is imposed by vacuum stability, $4 \lambda_s \lambda_h > \lambda_{hs}^2$.   
  Furthermore, in regions with a substantial $\lambda_s$, the  process $ss\rightarrow ssss$ becomes efficient and can lead to self--thermalization.
  We exclude these as well. 
   An additional Kaon physics constraint  $\theta < 10^{-4}$ applies for a very light $s$, $m_s < 200$ MeV. We find, however, that it is satisfied automatically. 
  
  Our numerical results are presented in Fig.\;\ref{planck-non-thermal-s}.
 The behaviour of the PLANCK curve can be understood as follows. The  factors that determine the neutrino abundance  are $Y_s$ and the decay branching fraction for 
 $s\rightarrow \nu\nu$. Consider first the regime $m_s \gg m_h$. In this case, $Y_s$ is determined by the fusion process $hh \rightarrow s$, whose rate is 
 proportional to $u^2 = M^2/\lambda^2$.  It terminates at temperatures of order $m_s \gg M$, so the $s$--yield scales simply as $M^2/\lambda^2$ with $M$ and $\lambda$.
 Now there are two options: $s$--decay can be dominated either by the sterile neutrino mode or by the SM channels.
  For $\Gamma (s\rightarrow \nu\nu) \gg \Gamma (s \rightarrow {\rm SM})$, the branching ratio 
  $ {\rm BR}(s\rightarrow \nu\nu)$ can be approximated by 1.
   Since $Y_\infty \propto 1/M$,
 the PLANCK line then satisfies $\lambda \propto M^{3/2}$. 
 In the opposite case  $\Gamma (s\rightarrow \nu\nu) \ll \Gamma (s \rightarrow {\rm SM})$, the branching ratio scales with $\lambda$ and $M$ as $  {\Gamma (s\rightarrow \nu\nu)  \over \Gamma 
 (s \rightarrow {\rm SM})} \propto   \lambda^2 / \theta^2 \propto \lambda^4/M^2$
 at $m_s\gg M$. This results in $\lambda \propto M^{-1/2}$. Thus, we have:
 \begin{eqnarray}
  m_s \gg m_h: && \nonumber\\
 &&          \lambda \propto M^{3/2}   ~~~{\rm for ~larger~}\lambda  \nonumber\\
 &&           \lambda \propto M^{-1/2} ~~{\rm for ~smaller~}\lambda
 \end{eqnarray}
 This scaling is observed in the  left panel of  Fig.\;\ref{planck-non-thermal-s}.
  
  For $m_s \ll m_h$, the $s$--abundance is dominated by $h\rightarrow ss$. If $s$ decays predominantly into neutrinos, the DM yield is independent of $\lambda$. Otherwise,
  it is proportional to $\lambda^4/M^2$. Thus, we get
  \begin{eqnarray}
  m_s \ll m_h: && \nonumber\\
 &&          M= {\rm const}   ~~{\rm for ~larger~}\lambda  \nonumber\\
 &&           \lambda \propto M^{1/4} ~~~~{\rm for ~smaller~}\lambda
 \end{eqnarray}
This behaviour is seen in the right panel of the figure. In both panels, DM is under-abundant to the left of (or below)  the PLANCK curve.

We see that quite large values of $\lambda$ up to $10^{-3}$ are consistent with all of the constraints. One may worry that the neutrinos would thermalize 
via $s \leftrightarrow \nu\nu$ at such a large coupling. However, the density of $s$ is much lower than its equilibrium value and this reaction does not increase the number of $s$--quanta, while $\nu\rightarrow \nu s$ is not allowed kinematically and $\nu\nu \rightarrow ss$ is suppressed. Thus, the system is not expected to thermalize.

  For a very light $s$, the BBN constraint on the lifetime of $s$ becomes significant: at small $\lambda$, it decays mostly into the photons and electrons which affect the abundance of light elements unless $\tau_s < 1$ sec. 
  
  Finally, we find that the mixing angle is very small in all the cases considered and its effects can be neglected.

   \subsubsection{On electroweak phase transition effects}
   
   The EW phase transition can have an important impact on the DM abundance. The Higgs mass reduction close to the transition opens up the  fusion channel
   $$ hh \rightarrow s$$
   even if this process is forbidden kinematically at other temperatures.
   It is operative if $2m_h(T\simeq T_c^v) < m_s $, while 
    its efficiency depends on the nature of the transition. (An analogous effect in a different setting was considered in \cite{Baker:2017zwx}.)
   
   In this work, we are interested in small couplings. Then, the electroweak phase transition corresponds either to a second order phase transition or a crossover.  In the former  case,
   the Higgs becomes massless at the critical temperature, while at the crossover it remains massive. Perturbative analysis is insufficient to distinguish the two: what appears as a second order transition typically corresponds to a crossover, as established by lattice simulations. 
   The full analysis of the singlet scalar extension is not yet available, although for a heavy singlet or EW triplet, the nature of the transition has been determined
   in \cite{Brauner:2016fla,Niemi:2018asa,Gould:2019qek}. The second order transition is found to occur in special cases, while a crossover is very common at weak coupling. This is to be contrasted with perturbative calculations (see e.g.\,\cite{Kurup:2017dzf}).
   Similar results are expected to apply in the light singlet or triplet  case.\footnote{We thank Lauri Niemi for sharing some of his results.}
    
    Although the Higgs does not turn massless at the crossover, its mass gets significantly reduced. In the SM, this reduction reaches an order of magnitude at the (pseudo-)critical temperature \cite{DOnofrio:2015gop} (see also earlier work \cite{Kajantie:1995kf,Kajantie:1996qd}). 
    Since we are mostly interested in very small Higgs portal couplings, the presence of the singlet is not expected to change the nature of the transition. Thus, we may assume 
    $m_h(T_c^v ) \sim 10 $ GeV as in the SM. 
    
       \begin{figure}[h!]
\begin{center}
\includegraphics[scale=0.345]{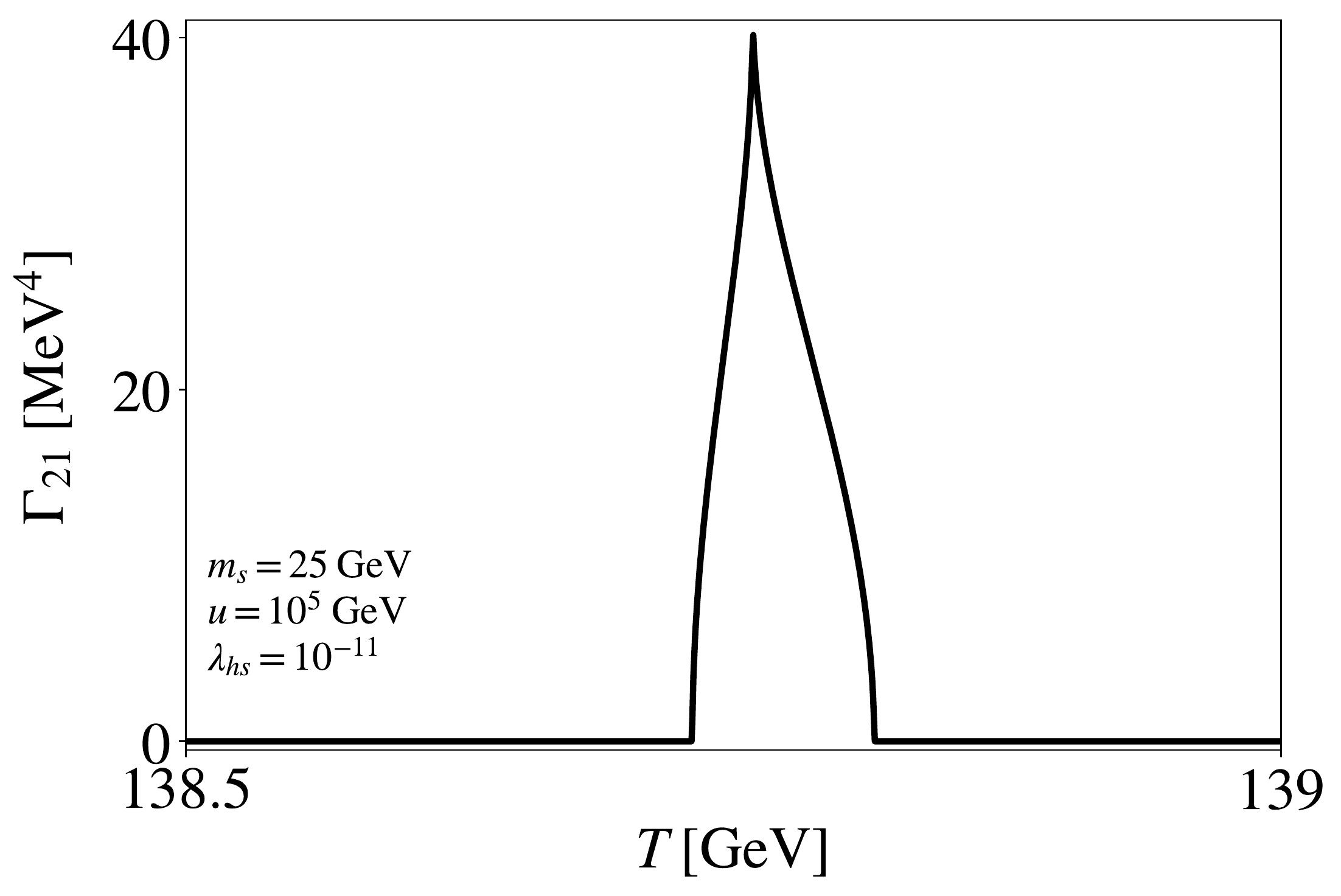}
\includegraphics[scale=0.345]{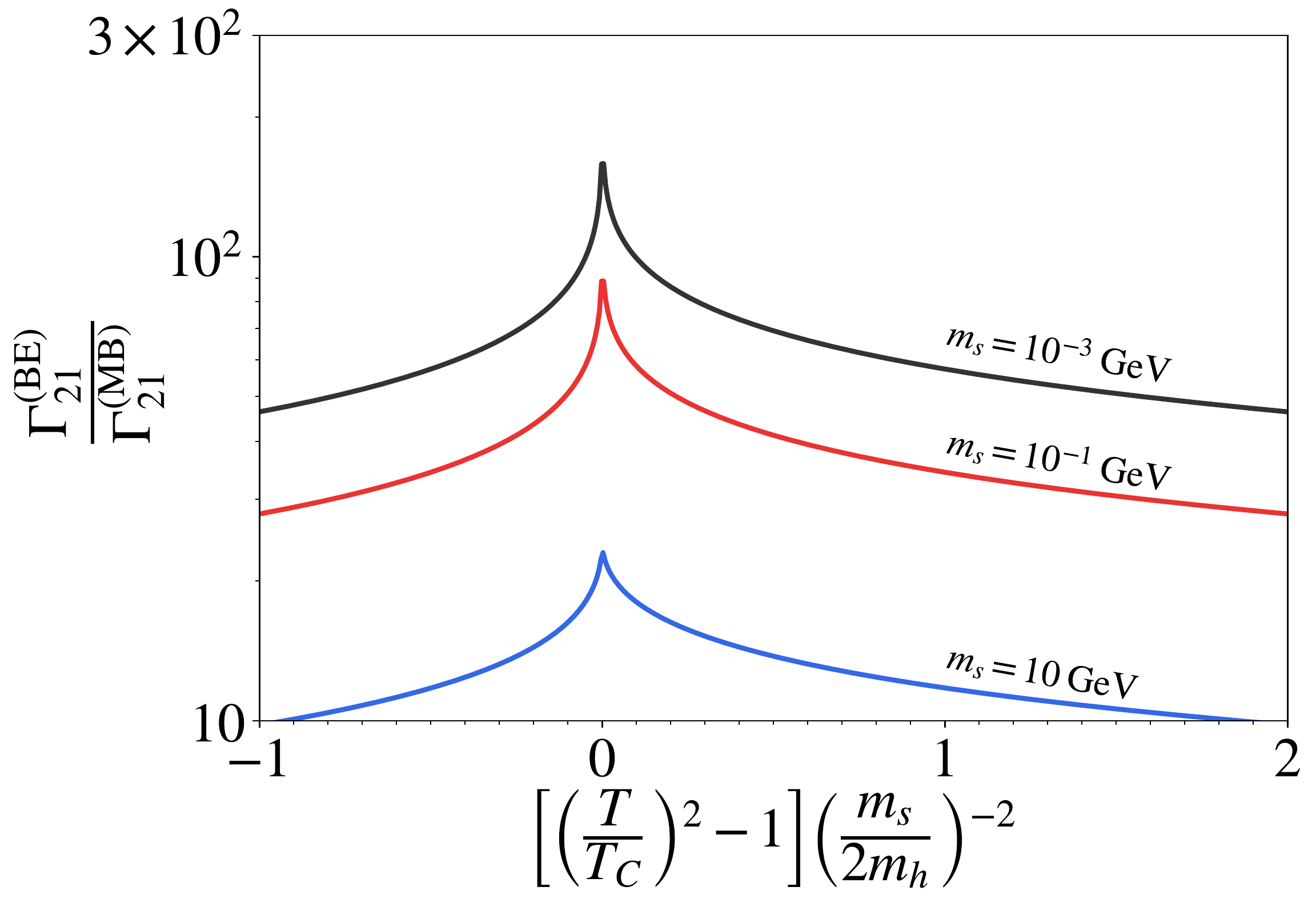}
 \end{center}
\caption{   {\it Left:}   Estimate of the $hh \rightarrow s$ reaction rate    at the EW crossover 
with $m_h(T_c)=10$ GeV.  The $s-h$ mixing 
is set to zero.  
{\it Right:} Bose--Einstein enhancement factor for $hh \rightarrow s$ at the 2d order EW phase transition.}
\label{phase-tr}
\end{figure}

    To estimate the efficiency of the fusion mode, let us consider a simplified case of zero $s-h$ mixing and employ a simple parametrization\footnote{
   This parametrization is inspired by the perturbative description of the 2d order phase transition, while it does not quite hold non--perturbatively. Nevertheless, it is acceptable for our estimate since the production is dominated by $T\simeq T_c$ where the Higgs mass is almost constant. } 
        \begin{eqnarray}
    && m_h^2 (T) = c (T^2-T_c^2)+m_h^2(T_c) ~~ ~{\rm for}~~T>T_c ~,\nonumber\\
     && m_h^2 (T) = 2c (T^2_c-T^2)+m_h^2(T_c) ~~ {\rm for}~~T<T_c ~,
    \end{eqnarray} 
   where $T_c \equiv T_c^v$ is the EW critical temperature and $c$ is a constant fixed by requiring $m_h(0)=125$ GeV.
   Taking a simple perturbative estimate for $T_c$, one can then calculate the fusion rate.
   The resulting $hh \rightarrow s$ rate for a representative parameter set is shown in Fig.\,\ref{phase-tr}, left panel.

 We find that this effect $alone$ can account for all of the observed dark matter. Although short, the fusion  is intense enough to produce numerous $s$--quanta which subsequently decay into sterile neutrinos. As one gets closer to the 2d order transition (at larger $\lambda_{hs}$), the Bose--Einstein enhancement becomes more pronounced. This is illustrated in
   Fig.\,\ref{phase-tr}, right panel. When 
    both $m_h(T_c)$ and $m_s$ are far smaller than the temperature, the Bose--Einstein 
    enhancement factor can reach  orders of magnitude.

The fusion mode can be more efficient than the decay $h \rightarrow ss$. Indeed, the fusion rate grows as $u^2$ which can be very large, while the decay rate remains constant for a fixed $m_s$. Thus, the thermalization constraints in Figs.\;\ref{lambda-hs-therm},\ref{lambda-s-therm} due to the fusion mode extend to $m_s<2m_h$ as well and can be more stringent then those due the decay, depending on $u$. However, in view of the uncertainties, we have not included these to be conservative.

  We note that our approximation   breaks down at $m_s \sim m_h$, i.e. when the mixing angle becomes significant.  As pointed out in \cite{Heeba:2018wtf}, the resonantly enhanced  $s-h$ mixing leads to additional scalar production. With present tools, it is however difficult to estimate its efficiency and we leave it for future work. We stress that the fusion mechanism considered here is intrinsically different and operative  for  small (and zero) mixing as long as $2 m_h(T) < m_s$.

\section{Conclusion}
\label{conclusion}

The lightest sterile neutrino is an attractive dark matter candidate. Although it is not stable, its longevity is guaranteed by its small mass and a small sterile--active mixing angle.
In this work, we explore the mass range up to 1 GeV. In this case, tiny mixing angles are necessary which one can  justify by a flavor--dependent  (neutrino parity) symmetry. 

We have focused on the scenario where the Majorana masses are entirely due to a VEV of a real scalar. This is enforced by a discrete lepton number symmetry, which is broken spontaneously by the scalar VEV. The scalar is then only allowed to couple to the SM quadratically  through the  Higgs portal.

Since the neutrinos can be very weakly coupled, the natural  (but generally not unique)   dark matter production mechanism is the freeze--in.
We have analyzed freeze--in production of sterile neutrinos ($\nu$) from the Higgs and singlet scalar ($s$) thermal bath   in different regimes. These are summarized in the following table.

 \begin{center}
\begin{tabular}{ |c|c|c| } 
 \hline
  & regime & dominant modes \\ 
 \hline
 thermal $s$ &  &  \\ 
 \hline
  & $m_s > 2M$ &  $s\rightarrow \nu\nu$ \\ 
   & $m_s < 2M$ &   $ss\rightarrow \nu\nu$, $h\rightarrow \nu\nu$, $s\rightarrow \nu\nu$  \\ 
   \hline
 non--thermal $s$ &  &  \\   
 \hline
 & heavy $s$ & $h\rightarrow \nu\nu$ \\
 & feebly coupled $s$ & $s\rightarrow \nu\nu$ \\
 \hline
\end{tabular}
\end{center}

 In all of these cases,   the observed DM relic density can be obtained. For the sterile neutrino mass range (1 keV, 1 GeV), we find that the requisite scalar--neutrino coupling varies
between $10^{-10}$ and $10^{-3}$.
 Our analysis takes into account the relativistic reaction rates with the Bose--Einstein distribution function, thermal masses and main effects of the phase transitions. All of these factors make an important impact on the final results. As byproducts, we have derived  relativistic rates for asymmetric reactions as well as  non--thermalization constraints on sterile neutrinos and the Higgs portal scalar. 
 
 We find a number of interesting effects which deserve further study. In particular, a light  scalar can be copiously produced close to the EW phase transition/crossover through the fusion mode $hh \rightarrow s$. Subsequent decay of the scalar into sterile neutrinos can account for all of the dark matter. However, the specifics of this mechanism require understanding non--perturbative dynamics close to the critical temperature.

The dark matter candidate studied here is long--lived. Its production mechanism is independent of the sterile--active mixing $\Theta$, hence there is vast parameter space  
$(\Theta, M)$ where dark matter decay can lead to an observable signal, 
e.g. in the form of monochromatic X- or gamma rays. The intensity of the signal  is 
 correlated with the dark matter density.\\

{\bf Acknowledgements.} OL is indebted to Mark Hindmarsh, Lauri Niemi and Aleksi Vuorinen for invaluable discussions.
VDR acknowledges financial support by the SEJI/2018/033 grant, funded by Generalitat Valenciana and partial support by the Spanish grants  FPA2017-85216-P and FPA2017-90566-REDC (Red Consolider MultiDark).
DK is supported  by the National Science Centre, Poland, research grant  No.  2015/18/A/ST2/00748. 
This work was made possible by Institut Pascal at Universit\'e
Paris-Saclay with the support of the P2I and SPU research departments and
the P2IO Laboratory of Excellence (program ``Investissements d'avenir''
ANR-11-IDEX-0003-01 Paris-Saclay and ANR-10-LABX-0038), as well as the
IPhT.  TT acknowledges funding from the Natural Sciences and Engineering Research Council of Canada (NSERC). 
Numerical computation in this work was carried out at the Yukawa Institute Computer Facility.

\appendix
\section{Leading thermal corrections}
In this Appendix, we summarize the most important thermal corrections to  the effective potential in our model.

The tree-level effective scalar potential, written in terms of the vevs $v, u$ reads
\begin{equation}
\label{pot}
V^0 =\frac{\lh}{4} v^4 + \frac{\lhs}{4} v^2 u^2 + \frac{\ls}{4} u^4 + \frac 12 \; \mu_h^2 \, v^2 + \frac 12 \mu_s^2 \, u^2 \,.
\end{equation}
The zero-temperature one-loop correction to effective potential is given by the Coleman-Weinberg correction~\cite{Coleman:1973jx}, which in the $\overline{MS}$ renormalisation scheme is
\begin{equation}
V^1  = \sum_\alpha \frac{n_\alpha}{64 \pi^2} m^4_\alpha(v,u)
\left(\log \frac{m^2_\alpha(v,u)}{Q^2} - C_\alpha
\right) \, .
\label{eq:V10T}
\end{equation}
Here  $\alpha$ runs over all dominant degrees of freedom: $t,~W,~Z,~G^{\pm,0}$ and $\chi_{1,2}$ (the mass eigenstates of the scalar fields $h$ and $s$).
The number of d.o.f.  $n_\alpha$ are  given by $n_t = -12,~ n_W = 6, ~n_Z = 3,~n_G = 3,~n_{\chi_{1,2}} =1$ (it includes a minus sign for fermions). 
$m^2_\alpha(v, u)$ are the field-dependent masses-squared,  $C_\alpha = 3/2 ~(5/6)$ for scalars (gauge bosons) and  $Q$ is the renormalisation scale. In our calculations, we take $Q$ to be  the particle masses in the vacuum at zero $T$. 
The field-dependent masses are: 

\begin{eqnarray}
\label{eq:fielddepmasses}
m_t^2(v,u) &=& y_t^2 \frac{v^2}{2} ~,\\
m_W^2(v,u) &=& g^2 \frac{v^2}{4}~,\\
m_Z^2(v,u) &=& (g^2+g^{\prime 2}) \frac{v^2}{4}~,\\
m_{G^0}^2(v,u) = m_{G^\pm}^2(v,u) &=& v^2 \lh + \frac{\lhs u^2}{2} + \mu_h^2~,\\
m_{\chi_{1,2}}^2(v,u) &=& v^2 \lh + \ls u^2 \pm \sqrt{v^4 \lh^2  + v^2 u^2 (\lhs^2 - 2 \lh \ls)+ \ls^2  u^4}.
\end{eqnarray}

The  temperature  effects are conveniently split into a one-loop temperature--dependent part $V^{1,T}$ and the ring corrections $V^{T}_{\rm ring}$~\cite{Dolan:1973qd, Carrington:1991hz}.
The former is  given by the one-loop thermal integral
\begin{equation}
 V^{1,T}(T) = \sum_\alpha \frac{n_\alpha T^4}{2 \pi^2} \mathcal{I}_{b,f}
\left(\frac{m_\alpha^2(v, u)}{T^2}\right)\;,
\label{eq:V1T}
\end{equation}
where  
\begin{equation}
\mathcal{I}_{b,f} \left(\frac{m_\alpha^2(v, u)}{T^2}\right) = \int_0^\infty dx~x^2 \log\left[1 \pm e^{-\sqrt{x^2+y^2}} \right] \, ,~y^2 = m^2_\alpha(v, u)/T^2 \;,
\label{eq:integralfb}
\end{equation}
with the plus (minus) sign for fermions (bosons), respectively. 
The ring contribution is   present only for bosons (gauge bosons, scalars and Goldstones):

\begin{eqnarray}
V^T_{_{\rm ring}} &=& -\frac{T}{12 \pi} \biggl\{ {\rm Tr}\left[(m_{gb}^2 + \Pi_{gb})^{3/2} - (m_{gb}^2)^{3/2}\right] + {\rm Tr}\left[(m_{\rm \chi}^2 + \Pi_{\rm \chi})^{3/2} - (m_{\rm \chi}^2)^{3/2}\right] \nonumber\\
&+& n_{G}\left[(m_{G}^2 + \Pi_{G})^{3/2} - (m_{G}^2)^{3/2}\right] \biggr\} \, ,
\label{eq:Vring}
\end{eqnarray}
where $m_{\chi}$ is the tree level scalar mass mixing matrix whose eigenstates are $\chi_{1,2}$. The squared mass mixing matrix for the electroweak gauge bosons is: 
\bea
m_{gb}^2 &=& \left(
                 \begin{array}{cccc}
                   \frac{g^2}{4}v^2 & 0 & 0 & 0 \\
                   0 & \frac{g^2}{4}v^2 & 0 & 0 \\
                   0 & 0 & \frac{g^2}{4}v^2 & -\frac{g g'}{4}v^2 \\
                   0 & 0 & -\frac{g g'}{4} v^2& \frac{g'^2}{4}v^2 \\
                 \end{array}
               \right) \, .
\label{eq:gbmassmatrix}
\eea

The $\Pi_i$ are the thermally corrected contributions to the masses~\cite{Carrington:1991hz,Ahriche:2007jp,Profumo:2007wc}:

\bea
\Pi_{gb} &=& \frac{11}{6}T^2 {\rm diag} \left(g^2, \;  g^2 , \; g^2, \; g'^2 \right) ~,\nonumber
\\ \nonumber
\\
\Pi_{\chi} &=&
\frac{1}{4}  T^2 {\rm diag} \left[ \left(\frac{3}{4} g^2 + \frac{1}{4} g'^2 + 2 \lh + y_t^2 + \frac{4 \lhs}{3} \right), \ls + \frac{1}{3} \lhs \right]  ~, \nonumber
\\ \nonumber
\\
\Pi_{G} &=&
\frac{1}{4}  T^2 \left(\frac{3}{4} g^2 + \frac{1}{4} g'^2 + 2\lh  + y_t^2 + \frac{4 \lhs}{3} \right) \,.\label{eq:Pithermalmasses}
\eea

The effective potential is then given by the sum of all of the above contributions:
\begin{equation}
V^T_{\rm eff}= V^0+V^1+V^{1,T}+V^T_{_{\rm ring}} \,.
\label{eq:effectiveV}
\end{equation}

In our analysis, we keep only the most important terms. We assume $V^1$ to be negligible at high temperatures and  use a  $1/T $  expansion of the   integrals in Eq.~\ref{eq:integralfb}  \cite{Dolan:1973qd}: 
\bea
\mathcal{I}_{b}\left(\frac{m}{T}\right) &\approx& -\frac{\pi^4}{45} + \frac{\pi^2}{12}\frac{m^2}{T^2} \,, \nonumber \\
\mathcal{I}_{f}\left(\frac{m}{T}\right) &\approx& \frac{7\pi^4}{360} - \frac{\pi^2}{24}\frac{m^2}{T^2} \,.
\label{eq:besselapprox}
\eea
It further proves convenient to use the expansion of the trace:
\begin{equation}
\left[(m_{i}^2 + \Pi_{i})^{3/2} - (m_{i}^2)^{3/2}\right] \approx \Pi_{i}^{3/2} + \frac{3}{2} \rm Tr \left[m_{i}^2 \sqrt{\Pi_{i}} \right] \,.
\end{equation}
Ignoring all field--independent terms, which shift the potential by a temperature dependent constant, we find that the ring corrections are of higher order in the couplings ($\sim g^3$) and can be neglected. The effective potential takes the form:
\begin{equation}
V^T_{\rm eff} = \frac{\lh}{4} v^4 + \frac{\lhs}{4} v^2 u^2 + \frac{\ls}{4} u^4 + \frac 12 \left[ (c_h T^2 + \mu_h^2) v^2 + (c_s T^2 + \mu_s^2) u^2  \right] \,,
\label{eq:Vefffinal}
\end{equation}
with $c_h=\frac{1}{4} \left(\frac{2 g^2}{4} + \frac{g^2 +g'^2}{4} +y_t^2 +2 \lh + \frac{\lhs}{6} \right)$ and $c_s=\frac{1}{4} \left(\ls + \frac{2}{3}\lhs \right)$.

\section{$s$ decay partial widths}

The real scalar $s$ interacts with the SM particles via its mixing with the Higgs. Its decay rates can be obtained from the Higgs ones  \cite{Djouadi:1997rp}  by including  the  factor $\sin^2 \theta$.  For $0.1~{\rm GeV} < m_s \lesssim 90$ GeV, we use the Higgs total decay width given in Refs.~\cite{Fradette:2017sdd,Winkler:2018qyg}. For masses $90~{\rm GeV} < m_s < 1000$ GeV, we   use the results of  Ref.~\cite{Dittmaier:2011ti}. Finally, for $m_s > 1$ TeV,  we scale the width up according to  $m_s^3$.

If $m_s < 2M$ and $M$ is in the keV range, $s$ will decay only to photons. 
In the calculation of the partial decay width  into photons, we follow~\cite{Spira:1995rr}:

\begin{equation}
\label{eq:Sdecwidthphph}
 \Gamma(s \to \gamma \gamma) = \frac{G_F \alpha^2 m_s^3 \sin^2 \theta}{128 \sqrt{2} \pi^3} \Bigg| \sum_f N^f_c Q_f^2 A_f(\tau_f) + N^W_c  Q_W^2 A_W(\tau_W)  \Bigg|^2,
\end{equation}
where   the sum runs over fermions and $W$ inside the loop. In this expression, $N^{f (W)}_c = 3 (1)$, $Q_i$ is the charge and $G_F$ is the Fermi coupling constant. 
We define the following mass ratio
\begin{equation}
 \tau_{\rm x} = \frac{m_s^2}{4 m_{\rm x}^2} \,
\end{equation}
and the loop functions 
\begin{align}
 A_f(\tau) &= 2 (\tau + (\tau -1) f(\tau))/\tau^2, \\
 A_W(\tau) &= -(2 \tau^2  + 3\tau  + 3 (2 \tau -1) f(\tau) )/\tau^2,
\end{align}
 with 
\begin{equation}
  f(\tau) = \begin{cases}
    \text{arcsin}^2 \sqrt{\tau} \hspace{1cm} &\text{for} \,\, \tau \le 1,\\
    -\frac{1}{4}\left(\log \frac{1+\sqrt{1-\tau^{-1}}}{1-\sqrt{1-\tau^{-1}}} -i\pi \right)^2 &\text{for} \,\, \tau > 1.
  \end{cases}
\end{equation}
Note that $\sin \theta$  depends on $m_s$:
\begin{equation}
\sin 2\theta =  \frac{M}{\lambda} \;\frac{2 \lhs v}{m_s^2 - m_h^2} \,.
\end{equation}

\begin{figure}[!hbt]
\begin{center}
\includegraphics[width=0.7\linewidth]{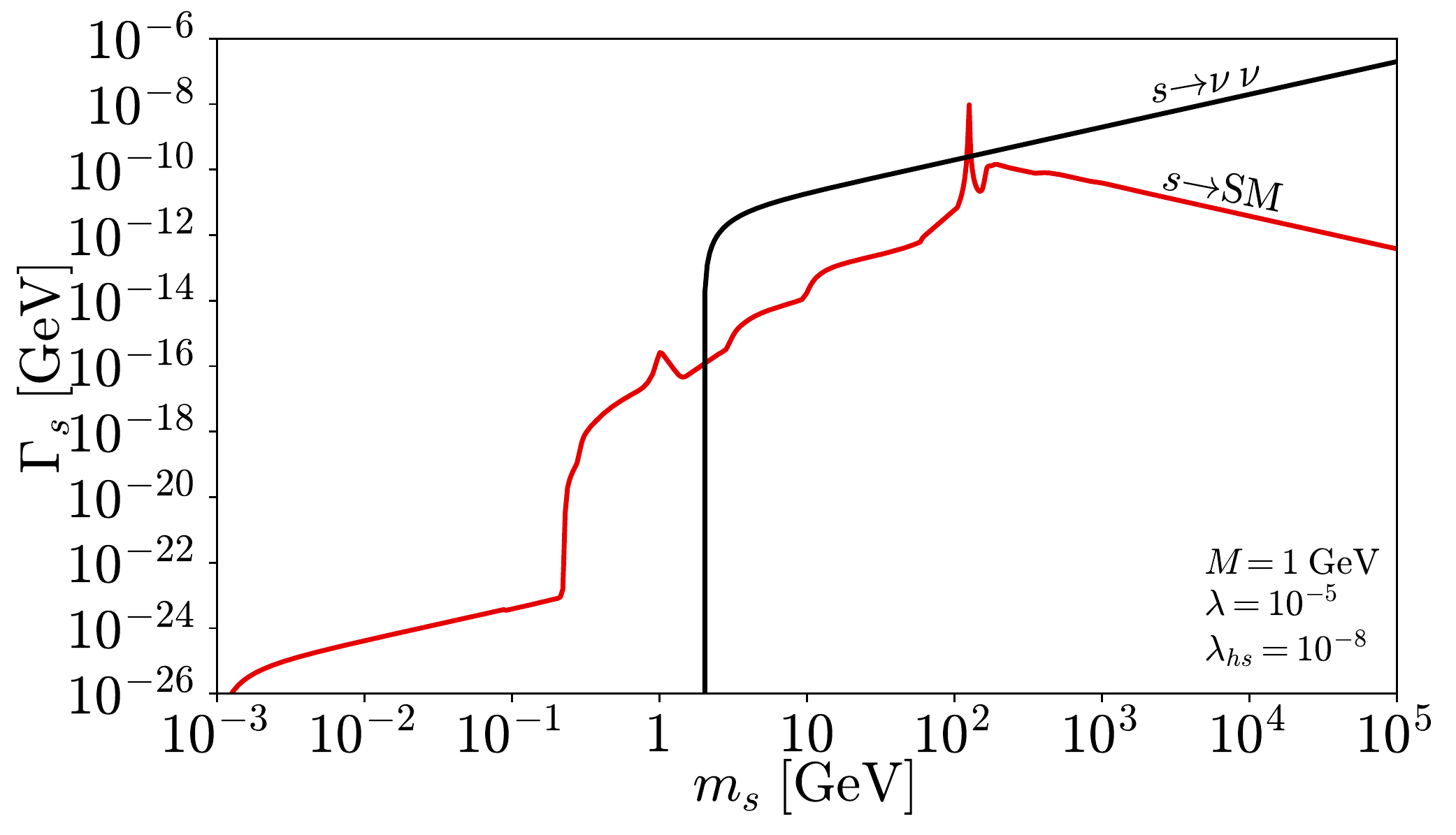}
\caption{The total SM decay width of $s$ and $\Gamma(s\rightarrow \nu\nu)$ for
 $\lambda = 10^{-5}$, $\lhs = 10^{-8}$ and $M = 1$ GeV.}
\label{fig:Stotalwidth}
\end{center}
\end{figure}

For heavier $m_s$, the scalar will also decay into other SM particles.
 Besides the SM channels, $s$ has another important decay mode $s\rightarrow \nu \nu$. 
 The corresponding decay width reads

\begin{equation}
\label{eq:Sdecwidthphph}
\Gamma(s \to \nu \nu) = \lambda^2 \frac{m_s}{16 \pi} \left(1 - 4 \frac{M^2}{m_s^2} \right)^{3/2}.
\end{equation}

  Fig.~\ref{fig:Stotalwidth}  shows the total SM decay width and $\Gamma(s \to \nu \nu) $  as a function of $m_s$ with other parameters fixed at some representative values. While the neutrino width grows with $m_s$,
  the SM decays get suppressed due to the decrease in the mixing angle $\theta \propto 1/m_s^2$. The spike in $\Gamma_s$ around $m_h\simeq m_s$ is due
  to the sharp increase in $\sin\theta$. In this region, our approximations are unreliable.

\bibliographystyle{utphys}
\providecommand{\href}[2]{#2}\begingroup\raggedright\endgroup

\end{document}